\documentclass[bibyear]{aa}  

\usepackage{graphicx}
\usepackage{txfonts}
\begin{document}

   \title{Global survey of star clusters in the Milky Way}

   \subtitle{IV. 63 new open clusters detected by proper motions}

   \author{R.-D. Scholz
          \inst{1}
          \and
          N.~V. Kharchenko\inst{2,3}
          \and
          A.~E. Piskunov\inst{2,4}
          \and
          S. R\"oser\inst{2}
          \and
          E. Schilbach\inst{2,5}
          }
\institute{
Leibniz-Institut f\"ur Astrophysik Potsdam (AIP), An der Sternwarte 16, D--14482
Potsdam, Germany\\
email: rdscholz@aip.de
\and
Astronomisches Rechen-Institut, Zentrum f\"ur Astronomie der Universit\"at
Heidelberg, M\"{o}nchhofstra\ss{}e 12-14, D--69120 Heidelberg, Germany
\and
Main Astronomical Observatory, 27 Academica Zabolotnogo Str., 03680 Kiev,
Ukraine
\and
Institute of Astronomy of the Russian Acad. Sci., 48 Pyatnitskaya Str., 109017
Moscow, Russia
\and
Max-Planck-Institut f\"ur Astronomie, K\"onigstuhl 17, D--69117 Heidelberg, Germany
             }

   \date{Received 14 April 2015 ; accepted 30 June 2015}

 
  \abstract
   {The global Milky Way Star Clusters (MWSC) survey carried 
    out by Kharchenko et al. provided new cluster membership
    lists and mean cluster parameters for nearly 80\% of 
    all previously known Galactic clusters. The MWSC
    data reduction pipeline involved the catalogue
    of positions and proper motions (PPMXL) on the 
    International Celestial Reference System (ICRS)
    and near-infrared photometry from the
    Two Micron All Sky Survey (2MASS).
    }
   {In their first extension to the MWSC, Schmeja et al. 
    applied photometric filters to the 2MASS to find new
    cluster candidates that were subsequently confirmed
    or rejected by the MWSC pipeline. To further extend 
    the MWSC census, particularly of nearby clusters,
    we aimed at discovering new clusters by conducting
    an almost global search in proper motion catalogues
    as a starting point.
    }
   {We first selected high-quality samples from the PPMXL
    and the Fourth US Naval Observatory CCD Astrograph 
    Catalog (UCAC4) for comparison and verification
    of the proper motions. For 441 circular proper motion 
    bins (radius 15~mas/yr) within $\pm$50~mas/yr, the sky outside a thin 
    Galactic plane zone ($|b|$$<$5$^{\circ}$) was binned in 
    small areas ('sky pixels') of 0.25$\times$0.25~deg$^2$.
    Sky pixels with enhanced numbers of stars with a certain
    common proper motion in both catalogues were considered 
    as cluster candidates. After visual inspection of the
    sky images, we 
    built an automated procedure that combined these
    representations of the sky for neighbouring proper motion subsamples
    after a background correction. The 692 compact cluster candidates detected 
    above a threshold that was equivalent to a 
    minimum of 12 to 130 cluster stars in 
    dependence on the Galactic latitude were then cross-checked
    with known star clusters and clusters of galaxies. 
    New candidates served as input for the MWSC pipeline.  
   }
   {About half of our candidates overlapped with known clusters
    (46 globular and 68 open clusters in the Galaxy, about 150 known 
    clusters of galaxies) or the Magellanic Clouds.
    About 10\% of our candidates turned out to be 63 new open clusters
    confirmed by the MWSC pipeline. They occupy predominantly the two 
    inner Galactic quadrants and have apparent sizes and numbers of 
    high-probable members slightly larger than those of the typically small 
    MWSC clusters, whereas their other parameters (ages, distances, tidal 
    radii) fall in the typical ranges. As our search aimed at finding compact 
    clusters, we did not find new very nearby (extended) clusters, and the 
    mean total proper motion of the new 63 clusters is
    with 6.3~mas/yr similar to the MWSC average (5.5~mas/yr).
    Only four new clusters have mean proper motions between 10~mas/yr and 
    our observed maximum of about 13~mas/yr.}
   {}


\titlerunning{Global survey of star clusters in the Milky Way IV.}
\authorrunning{R.-D. Scholz et al.} 

   \maketitle
%

\section{Motivation}\label{Sect_motiv}

Star clusters are considered as building blocks of the Galaxy.
Their study is important for our understanding of star formation
and the history of the Milky Way. They are possibly the birth 
places of all stars (Lada et al.~\cite{lada93}). Our sun was probably 
born in a star cluster that dissolved long time ago; and searches are
ongoing to find the lost solar siblings (see e.g. 
Portegies Zwart~\cite{portegies09};
Mishurov \& Acharova~\cite{mishurov11};
Bobylev \& Bajkova~\cite{bobylev14};
Batista et al.~\cite{batista14}).
A complete (volume-limited) census of the current population of 
Galactic star clusters is difficult to obtain
because of gas and dust clouds disturbing our view 
towards the 
Galactic plane. 
But at least in the solar neighbourhood, we should aim at
completeness
for investigating all the different
aspects related to Galactic star clusters. Systematic searches 
for missing nearby star clusters are necessary to fill the
open cluster census; and new large star catalogues can be used for 
this purpose.

Over the last decade, we have directed our efforts to improve
the membership and parameters of 
Galactic star clusters 
with a global
approach and making use of uniform all-sky data. As a result, we first 
presented combined photometric/proper motion membership probabilities 
(Kharchenko et al.~\cite{kharchenko04}) as well as cluster sizes, 
distances, colour-excesses, mean proper motions, and ages
(Kharchenko et al.~\cite{kharchenko05a}) for 520 open clusters.
This first cluster survey was based on the all-sky compiled catalogue of 
2.5 million stars (ASCC-2.5; Kharchenko~\cite{kharchenko01})
providing optical ($BV$) photometry and proper motions. The
ASCC-2.5 with its relatively bright stars served also as input for
membership analysis and the determination of the same set of cluster 
parameters as provided for the first 520 clusters
of another 130, including 109 newly discovered, open clusters
(Kharchenko et al.~\cite{kharchenko05b}). The resulting 
Catalogue of Open Cluster Data (COCD) included 650 clusters.

Additional cluster parameters were later provided:
radial velocities for a subsample of the 650 COCD clusters studied with 
the ASCC-2.5 and for additional fainter clusters
(Kharchenko et al.~\cite{kharchenko07}, Conrad et al.~\cite{conrad14}),
tidal radii and masses of all 650 open clusters in the COCD
(Piskunov et al.~\cite{piskunov08}), 
cluster shape parameters
(Kharchenko et al.~\cite{kharchenko09a}),
and integrated magnitudes and colours in optical ($BV$) and 
near-infrared ($JHK_s$) passbands together with luminosity functions
of all 650 clusters
(Kharchenko et al.~\cite{kharchenko09b}).

The COCD data have been extensively used by our group and by
many other authors in numerous follow-up studies. Among the topics of
these investigations are:
the Galactic open cluster population including spatial/kinematical 
substructures and cluster complexes
(Piskunov et al.~\cite{piskunov06}; Zhu~\cite{zhu09}; Wu et al.~\cite{wu09};
Elias et al.~\cite{elias09}; Vande Putte et al.~\cite{vandeputte10}; 
Gozha et al.~\cite{gozha12}; Morales et al.~\cite{morales13}),
the disruption of star clusters
(Lammers et al.~\cite{lammers05}; Gieles \& Bastian~\cite{gieles08}),
massive field stars and neutron stars 
(Posselt et al.~\cite{posselt07}; Schilbach \& R{\"o}ser~\cite{schilbach08}; 
Gvaramadze \& Bomans~\cite{gvaramadze08}; Hubrig et al.~\cite{hubrig11};
Tetzlaff et al.~\cite{tetzlaff14}),
Cepheids 
(Anderson et al.~\cite{anderson13}; Chen et al.~\cite{chen15}),
and binaries in open clusters
(de Grijs et al.~\cite{degrijs08}; Guerrero et al.~\cite{guerrero14}).

A much more complete cluster survey based on about 470 million objects 
with near-infrared ($JHK_s$) photometry from the Two Micron All Sky Survey
(2MASS; Skrutskie et al.~\cite{skrutskie06}) supplemented by proper motions
from the PPMXL catalogue (R\"oser, Demleitner \& Schilbach~\cite{roeser10})
was started by Kharchenko et al.~(\cite{kharchenko12}) (hereafter paper I.)
This Milky Way Star Clusters (MWSC) survey of all previously known
clusters was completed by Kharchenko et al.~(\cite{kharchenko13}) (paper II). 
For 3006 out of 3784 clusters 
mentioned in
the literature, photometric/proper motion membership probabilities  
and cluster parameters (cluster centre, apparent size, distance, mean   
proper motion, colour excess, and age) were determined with the new dedicated 
data-reduction pipeline described in paper I. 
Tidal radii were also derived for most of these 
clusters, whereas radial velocities were estimated for about 30\%.
While this MWSC sample may be considered as almost complete within a distance 
of 1.8~kpc from the Sun, 
we found evidence 
for possibly missing relatively old ($\log t \gtrsim 9$) 
and nearby (within about 1~kpc) clusters (paper II).

Of course, the new largely extended stellar database raised high hopes
for new cluster discoveries, too. However,
to search for new clusters in the 2MASS/PPMXL data using the 
pipeline described in paper I,
we first need some candidates or ''seeds''. In our first cluster search,
the COCD extension based on the ASCC-2.5, a few hundred bright stars with 
available spectral classification, previously not known as cluster members,
could be selected as cluster ''seeds'' (Kharchenko et al.~\cite{kharchenko05b}).
With the now available much larger stellar database, an obvious choice
of ''seeds'' would be stellar density enhancements in small sky areas.
The 2MASS alone was already used to look for such density enhancements
in the 
Galactic 
plane ($|b|$$<$20$^\circ$) by
Froebrich et al.~(\cite{froebrich07}), and their cluster candidates
were included in the above mentioned input list of 3784 known clusters.
A logical next step was to check the 2MASS data at higher Galactic
latitudes. This was recently performed by Schmeja et al.~(\cite{schmeja13}) 
(paper III) as a first extension of the MWSC survey. They 
discovered 
139 new, 
mainly old, open clusters within 
about 3~kpc by applying filters in the 2MASS near-infrared colour-magnitude 
space and applying the data reduction pipeline described in paper I 
in the sky regions of their candidates.

Members of a star cluster are expected not only to concentrate in
a small sky area but also to share a common proper motion. Therefore,
the combination of both effects seemed to be another
promising cluster search method. 
It should be in particular sensitive for nearby objects, as the mean 
cluster proper motions typically increase with smaller distances.
Here, we present the results of a first nearly all-sky search,
excluding only the immediate Galactic plane within $|b|$$<$5$^\circ$, for new
star clusters primarily based on the proper motions and positions of stars 
and involving the same large catalogues (PPMXL/2MASS) as used
in papers I, II, and III. This is the second extension of the 
MWSC survey.

%
\section{Quality check of global proper motion catalogues}\label{Sect_qual}

To be consistent with the MWSC survey (papers I, II, and III),
our cluster search, membership and parameter determination were 
based on proper motions from the PPMXL catalogue and 
its improved subset 2MAst (2MASS with astrometry) described in paper I.
In addition, we used the proper motions from the Fourth US Naval Observatory 
CCD Astrograph Catalog (UCAC4; Zacharias et al.~\cite{zacharias13}), but
only for the verification of our cluster candidates. Both 2MAst and UCAC4 are
accompanied by accurate near-infrared photometry and quality flags
from 2MASS.  Below we outline our 
selection criteria for the best data extracted from these two
catalogues, a comparison of these data sets and an investigation
of their systematic differences that may affect
a cluster search by proper motion.

   \begin{figure}
   \centering
   \includegraphics[width=8.3cm]{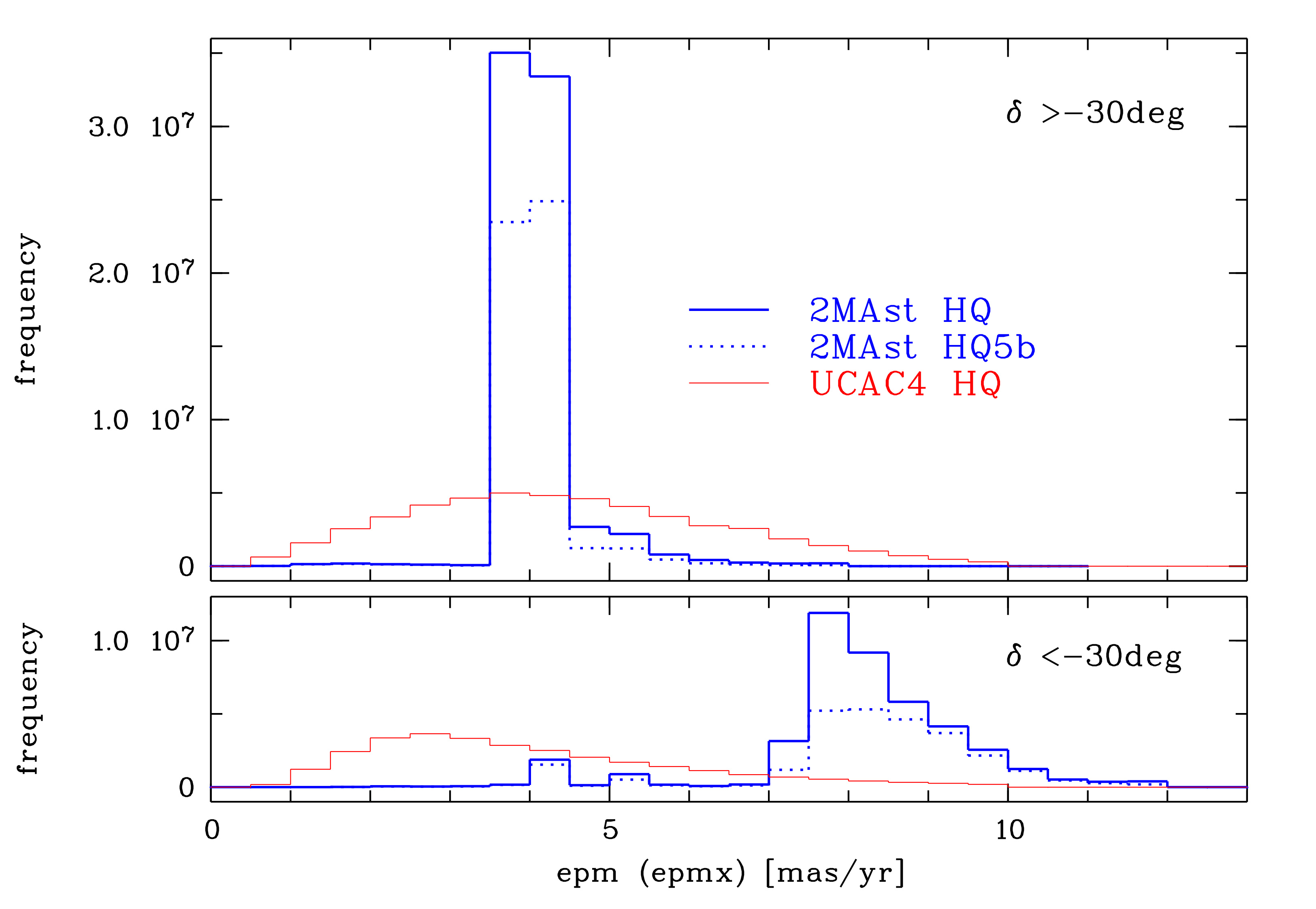}
      \caption{Distribution of proper motion errors (epm)
               in the chosen 2MAst\,HQ (blue solid line),
               2MAst\,HQ5b (blue dotted line),
               and UCAC4\,HQ (red thin line, here we show the proper motion
               errors in right ascension) samples.
               \textbf{Top:} north of $\delta$$=$$-$30$^{\circ}$,
               \textbf{Bottom:} south of $\delta$$=$$-$30$^{\circ}$.
              }
         \label{Fepm}
   \end{figure}

\subsection{Selection of PPMXL (2MAst) objects}\label{SubS_ppmxl}

The PPMXL catalogue of 
R\"oser, Demleitner \& Schilbach~(\cite{roeser10}) gives positions
and proper motions in the International Celestial Reference System (ICRS)
for about 900 million objects. As it was 
based on
the United States Naval Observatory (USNO) B1.0 catalogue
(Monet et al.~\cite{monet03}), the 
main source for determining the proper motions were 
measurements of 
photographic Schmidt plates. In certain sky regions, where these Schmidt 
plates overlap, the PPMXL contains multiple entries for the same objects,
which would be confusing for a cluster search. However, the merger of 2MASS
and PPMXL,
the 2MAst catalogue (see paper I), lists the averaged proper motions for
these objects and is therefore more appropriate for this task. It
contains proper motions and accurate 2MASS photometry for about
399 million objects. As shown in Fig.~9 of R\"oser, 
Demleitner \& Schilbach~(\cite{roeser10}), the excluded PPMXL objects with
lacking 2MASS measurement constitute the low-quality tail of the 
error distribution of the PPMXL proper motions starting at about 5~mas/yr 
and reaching up to 30~mas/yr.

To select a high-quality (HQ) subsample from the 2MAst catalogue, we
applied the following selection criteria:

\begin{enumerate}
\item 2MASS $JHK_s$ photometric quality 'AAA' 
\item Exactly one matching
object within 5~arcsec in USNO A2.0 (Monet et al.~\cite{monet98})
or Tycho-2 (H{\o}g et al.~\cite{hog00}) catalogues
\item More than three observations used for the proper motion
\item Proper motion differences (if available) between 
PPMX (R{\"o}ser et al.~\cite{roeser08}) and PPMXL
$<$6~mas/yr 
\item Proper motion errors $<$8~mas/yr at 
declinations $\delta$$>$$-$30$^{\circ}$ and $<$12~mas/yr at           
declinations $\delta$$<$$-$30$^{\circ}$
\end{enumerate}

The first two conditions aimed at minimising the influence of
extended (non-stellar) and overlapping images. 
These may be resolved to a different degree
on the first- and second-epoch Schmidt plates used in various versions of 
the USNO catalogues and lead to false proper motions in the PPMXL.
With the last three 
conditions we excluded uncertain proper motion measurements, where
the different PPMXL proper motion quality north and south of 
$\delta$$=$$-$30$^{\circ}$ 
(see Fig.~9 of R\"oser, Demleitner \& Schilbach~\cite{roeser10}) 
was taken into account. 
We only allowed for relatively small proper motion differences
between PPMX and PPMXL, as the PPMX was used in the construction of the
PPMXL and also based in part on Schmidt plates and 2MASS data.
With the above conditions
we selected about 119 million objects for this 2MAst\,HQ sample
(about 30\% of the whole 2MAst). Their proper motion error distribution
is shown in Fig.~\ref{Fepm}.
 
Because of the stronger image crowding on the Schmidt plates covering the 
Galactic plane, the PPMXL proper motions of both cluster and field stars 
become more unreliable with low Galactic latitudes. Therefore, we further 
excluded the Galactic latitude zone within $|b|$$<$5$^{\circ}$ and arrived 
at about 79.2 million objects. 

   \begin{figure}
   \centering
   \includegraphics[width=9.3cm]{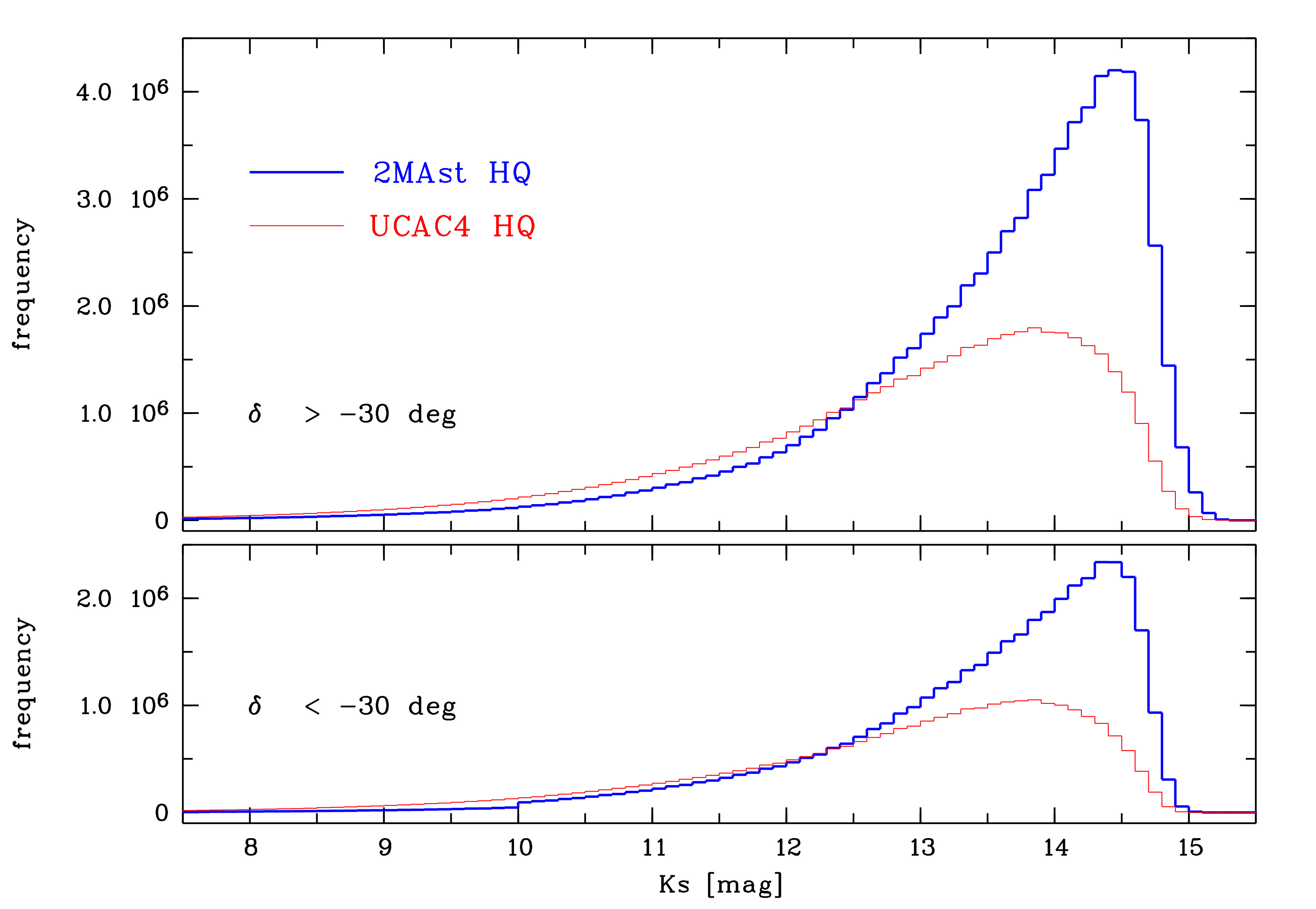}
      \caption{Distribution of 2MASS $K_s$ magnitudes
               in the chosen all-sky samples 2MAst\,HQ (blue solid line)
               and UCAC4\,HQ (red thin line).
               \textbf{Top:} north of $\delta$$=$$-$30$^{\circ}$,
               \textbf{Bottom:} south of $\delta$$=$$-$30$^{\circ}$.
              }
         \label{FKs}
   \end{figure}

\subsection{Selection of UCAC4 objects}\label{SubS_ucac4}

The UCAC4 (Zacharias et al.\cite{zacharias13}) provides proper motions
on the ICRS for about 105 million objects and also includes 2MASS photometry. 
UCAC4, unlike the previous version UCAC3, does not involve any 
Schmidt plate data 
in the
determined proper motions. Therefore, for the majority of fainter objects,
the UCAC4 proper motions are 
practically independent 
from those given in 
the PPMXL. Our HQ selection criteria applied to the UCAC4 were:

\begin{enumerate}
\item Combined 2MASS contamination and photometric flags $=$5
(no artefacts/contamination and quality 'AAA')
\item Combined UCAC4 double star flag $=$0
\item UCAC4 object flag $<$2
\item Proper motion errors $<$10~mas/yr
\end{enumerate}

The first two conditions are equivalent to the first two 2MAst conditions.
However, because of the higher spatial resolution of the observations 
used for determining UCAC4 proper motions compared to the Schmidt plates
used in 2MAst, the second condition is an even stronger 
constraint. Concerning the UCAC4 object flag (third condition), we
initially 
tried to use only zero values ('good, clean star, no known problem'), 
but noticed 
large areas of sky in a declination zone between $+$10$^{\circ}$
and $+$20$^{\circ}$, where 
for some unknown reason this flag was
always set to 1 ('near overexposed star'). 
Therefore, we 
finally
allowed it to be equal to 
zero or 1 all over the sky.
Finally, by the last condition of 
proper motion errors smaller than 10~mas/yr, we
decided for 
a value in between the two 
chosen 
limits in the 2MAst. As can be seen in Fig.~24 of
Zacharias et al.~(\cite{zacharias13}), the tail  of the error distribution 
of UCAC4 proper motions (10-35~mas/yr) contains less than 5\% of all objects. 
The final number of objects in our 
chosen UCAC4\,HQ sample was about 79.0 million
(about 75\% of all UCAC4 objects with proper motions). 
We show the
proper motion error 
distribution of the UCAC4\,HQ sample in Fig.~\ref{Fepm}.

   \begin{figure}
   \centering
   \includegraphics[width=9.3cm]{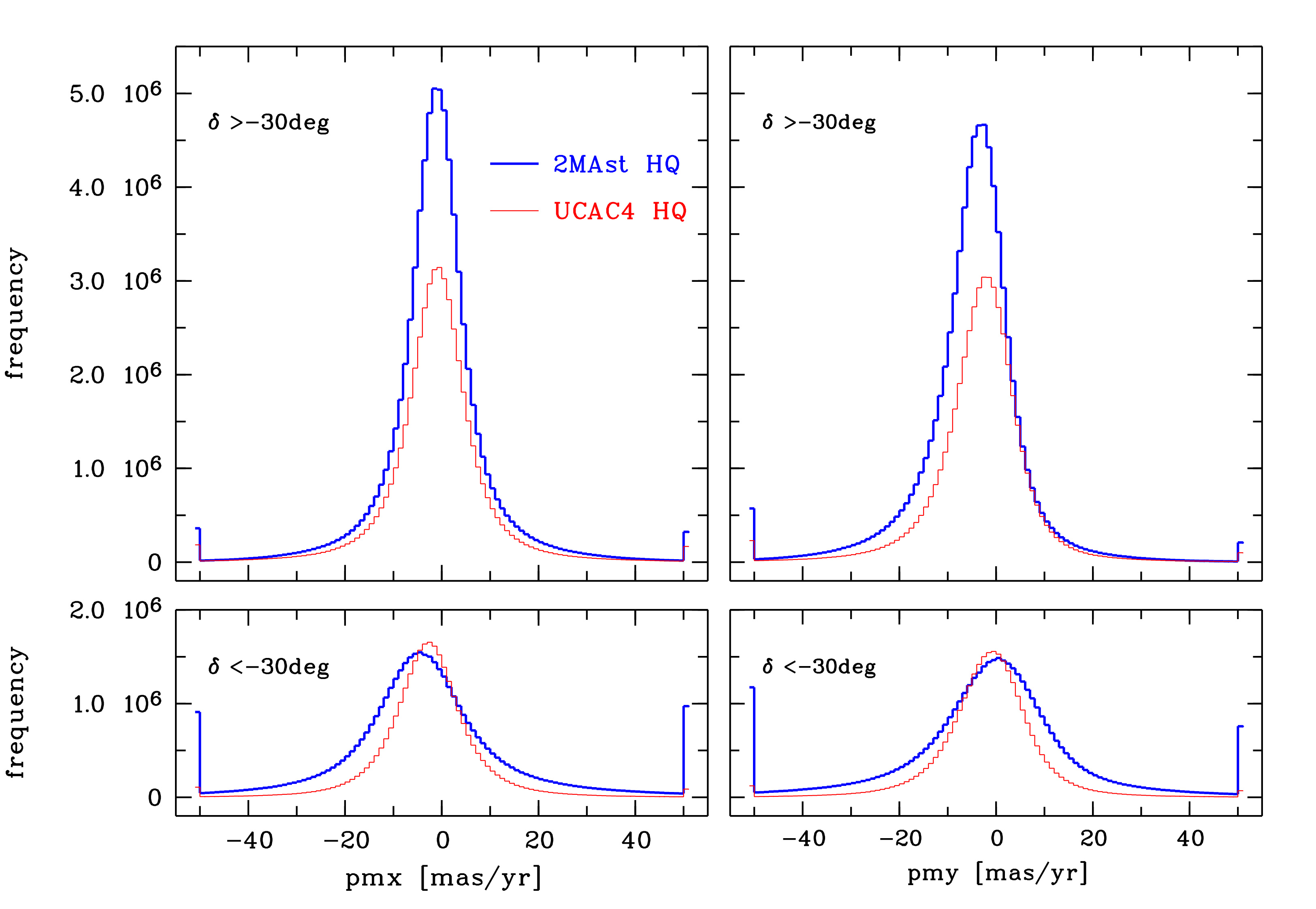}
      \caption{Distribution of proper motions in right ascension
               (pmx = $\mu_{\alpha}\cos{\delta}$)
               and declination (pmy = $\mu_{\delta}$)
               in the chosen all-sky samples 2MAst\,HQ (blue solid line)
               and UCAC4\,HQ (red thin line).
               \textbf{Top:} north of $\delta$$=$$-$30$^{\circ}$,
               \textbf{Bottom:} south of $\delta$$=$$-$30$^{\circ}$.
               The number of objects in the long tails of the distributions
               extending outside the shown range of $-$50...$+$50~mas/yr
               are indicated at the histogram edges.
              }
         \label{Fpm}
   \end{figure}

\subsection{Comparison of 2MAst\,HQ and UCAC4\,HQ}\label{SubS_comp2mastucac4}

As the PPMXL has a much fainter optical magnitude limit than the UCAC4, our 
chosen 
2MAst\,HQ and UCAC4\,HQ samples containing only the best-measured
objects with 2MASS counterparts are expected to reach different limiting
magnitudes in the near-infrared, too. This is confirmed by Fig.~\ref{FKs},
where we see that both samples include objects
up to 14th magnitude, but the 2MAst\,HQ distribution peaks
at $K_s$$\approx$14.5, more than half a magnitude 
fainter than 
in case of UCAC4\,HQ. The number of faint objects ($K_s$$>$12.5) in
2MAst\,HQ 
is nearly two times larger than in UCAC4\,HQ.
On the other hand, there are 
sightly more bright stars
($K_s$$<$12.5) in UCAC4\,HQ compared to 2MAst\,HQ. 

The distributions 
of proper motion errors in our 
chosen 
2MAst\,HQ and UCAC4\,HQ samples 
look 
very different (Fig.~\ref{Fepm}). Both
samples have their peak at about 4~mas/yr, if the sky
north of $\delta$$=$$-$30$^{\circ}$ is considered. However, the UCAC4\,HQ
errors occupy the whole range between about 1~mas/yr and our 
chosen
limit of 10~mas/yr, whereas the 2MAst\,HQ distribution is very sharp and
contains about 90\% of all objects in the two central bins (3.5-4.5~mas/yr).
This follows from the observational history as stated 
in R\"oser, Demleitner \& Schilbach~(\cite{roeser10}). In particular, this
is seen in their Fig.~9. As the typical epoch
differences of the Schmidt plates in the southern sky are much smaller,
the 2MAst\,HQ histogram changes dramatically
south of $\delta$$=$$-$30$^{\circ}$, where the majority of objects
exhibits errors between 7 and 10~mas/yr, whereas the UCAC4\,HQ peak shifts
to an even lower value at about 3~mas/yr.

We also checked 
the overall proper motion distributions of our 
chosen
HQ samples (Fig.~\ref{Fpm}). While we did not expect normal
distributions for the proper motions of all stars over large parts of the
sky, we considered the different widths of the distributions, and especially
the numbers of outliers (high proper motion stars) as possible
indicators of the reliability of the proper motions. By our HQ selection
criteria we successfully excluded
the vast majority of wrong high proper motion stars in the PPMXL catalogue
that stem from the B1.0 catalogue. R\"oser, Demleitner \&
Schilbach~(\cite{roeser10}) mentioned e.g. 
about 24.5 million PPMXL objects 
with apparent proper motions larger than 150~mas/yr in the northern hemisphere.
The 2MAst\,HQ histogram edges in the top and bottom panels of Fig.~\ref{Fpm} 
show only 
about 0.8 and 1.9 million objects 
with proper motions larger than 
50~mas/yr, respectively. However, compared to the UCAC4\,HQ histogram edges, 
these numbers are more than two times larger at $\delta$$>$$-$30$^{\circ}$ and
even about ten times larger at $\delta$$<$$-$30$^{\circ}$. The location of
the peak and the width of the pmx 
(we use pmx and pmy equivalent 
to $\mu_{\alpha}\cos{\delta}$ and $\mu_{\delta}$)
distributions of 2MAst\,HQ and UCAC4\,HQ
(with formally measured standard deviations of about 13~mas/yr and 11~mas/yr,
respectively) are in good
agreement only at $\delta$$>$$-$30$^{\circ}$. The peak locations in the other
three panels of Fig.~\ref{Fpm} differ by a few mas/yr, which may be a first
indication of systematic proper motion differences between the two catalogues,
although we should keep in mind the different sizes and magnitude distributions
(Fig.~\ref{FKs}) of the 2MAst\,HQ and UCAC4\,HQ samples. The much lower proper
motion quality of the 2MAst\,HQ at $\delta$$<$$-$30$^{\circ}$ 
(Fig.~\ref{Fpm})
leads to about
two times broader proper motion (both pmx and pmy) distributions (standard
deviations of about 24~mas/yr) compared to the UCAC4\,HQ (about 11~mas/yr).

   \begin{figure}
   \centering
   \includegraphics[width=9.3cm]{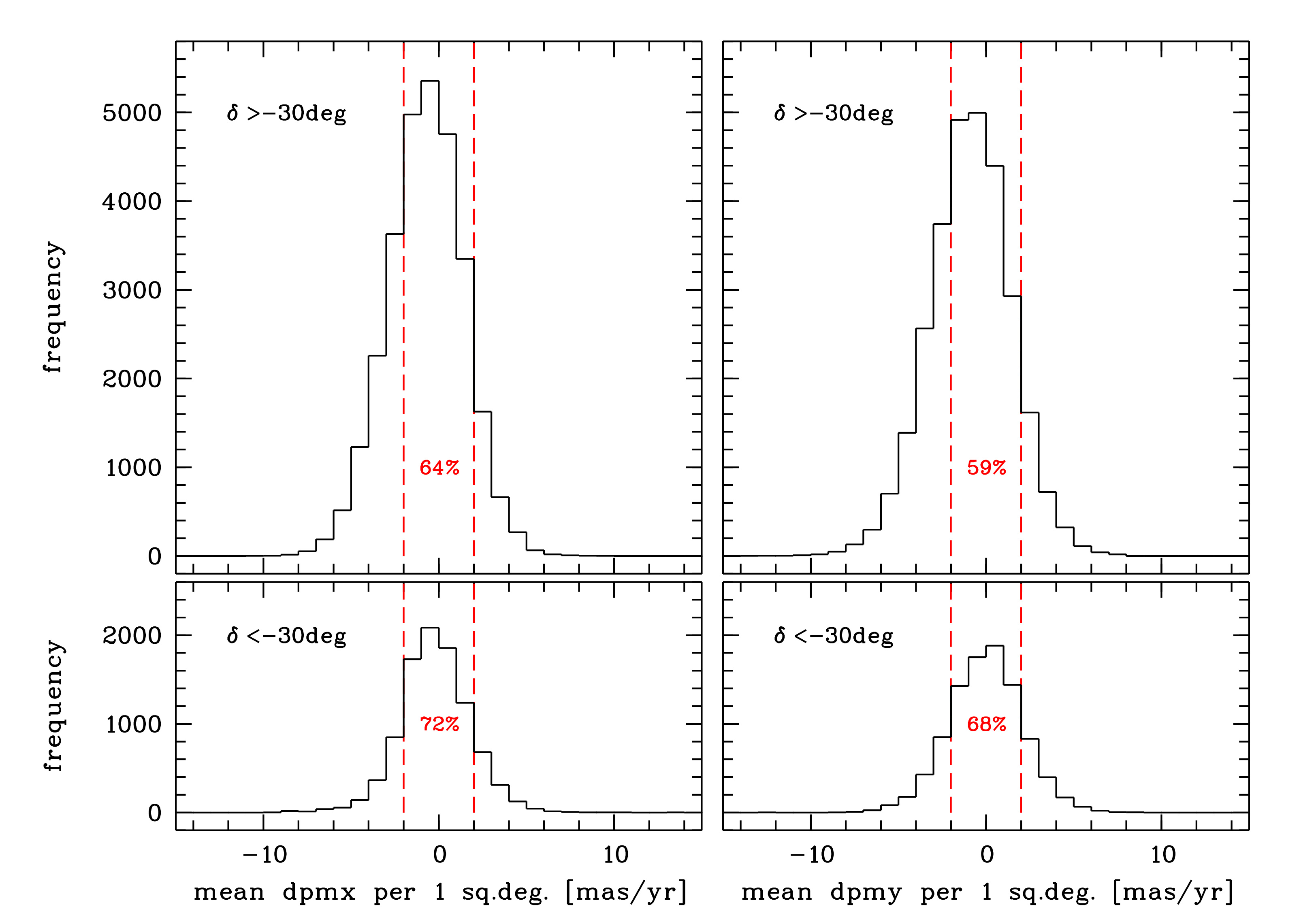}
      \caption{Histograms of mean proper motion differences between
               2MAst\,HQ5b and UCAC4\,HQ in 1~deg$^2$ areas of the sky
               in right ascension (dpmx) and declination (dpmy)
               \textbf{Top:} north of $\delta$$=$$-$30$^{\circ}$,
               \textbf{Bottom:} south of $\delta$$=$$-$30$^{\circ}$.
               The Galactic plane zone ($|b|$$<$5$^{\circ}$) was excluded.
               The outliers in each 1~deg$^2$ area were excluded by a
               3-$\sigma$ clipping. The fraction of areas with
               differences of less than 2~mas/yr is indicated
               in each panel.
              }
         \label{Fdpmsqb1his}
   \end{figure}

   \begin{figure*}
   \includegraphics[width=9.1cm]{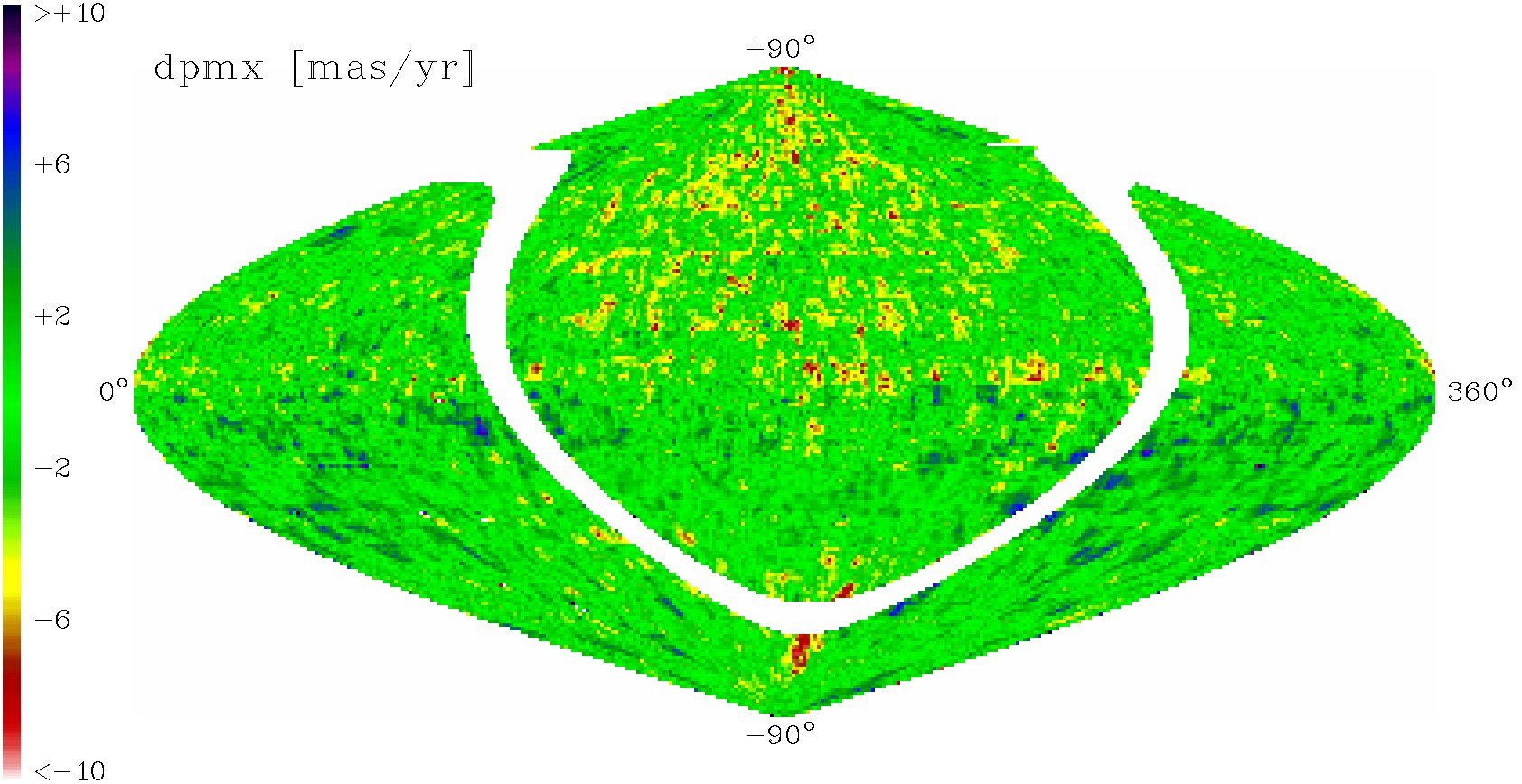}
   \includegraphics[width=9.1cm]{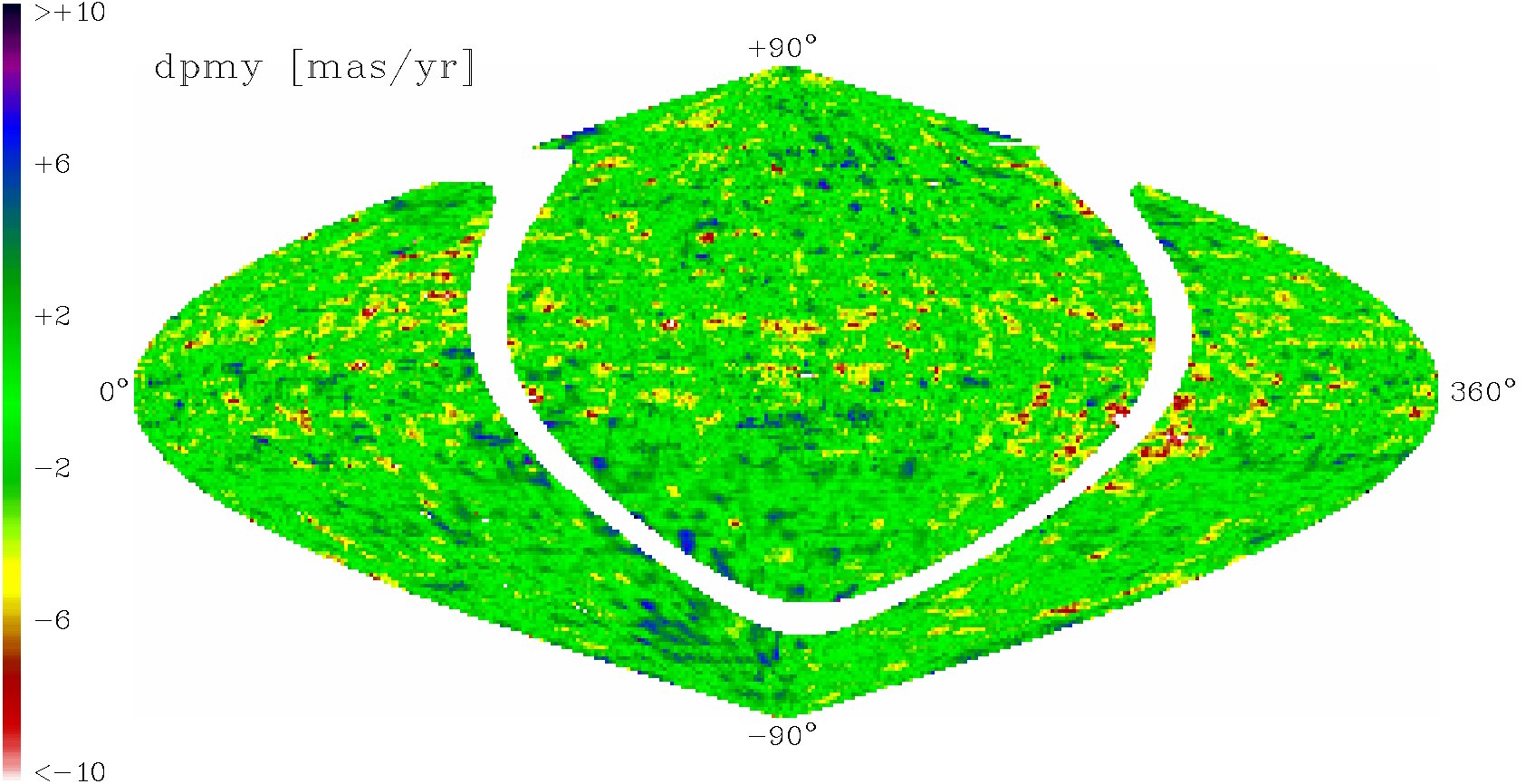}
      \caption{Distribution of mean proper motion differences
               \textbf{Left:}
               in $\mu_{\alpha}\cos{\delta}$ (dpmx) and
               \textbf{Right:}
               in $\mu_{\delta}$ (dpmy) between
               2MAst\,HQ and UCAC4\,HQ over the sky
               (projection of $\alpha\cos{\delta},\delta$).
               Each pixel represents an area of $\sim$1~deg$^2$.
              }
         \label{Fdpmsqb1}
   \end{figure*}

To exclude the effect of different sample sizes and magnitude distributions
of 2MAst\,HQ and UCAC4\,HQ, and to investigate systematic proper motion
differences as a function of the position in the sky, we identified a common 
2MAst/UCAC4\,HQb5 sample. Over the full sky except for the Galactic plane zone 
with $|b|$$<$5$^{\circ}$, that is in an area of $\sim$37700~deg$^2$, we
found 
about 44.0 million objects, 
which met our quality constraints, by 
using a matching radius of 1~arcsec. R\"oser, Demleitner \&
Schilbach~(\cite{roeser10}) estimated the systematic proper motion
errors in small sky areas on a 1$^{\circ}$ scale to be at least 1-2~mas/yr
for the PPMXL, but also for the UCAC series. We binned our common 
2MAst/UCAC4\,HQb5 sample in $\sim$1~deg$^2$ areas of the sky using
a simple $\alpha\cos{\delta},\delta$ projection and determined
the mean proper motion differences after applying a 3-$\sigma$ clipping
in each bin. The results are shown in Figs.~\ref{Fdpmsqb1his} 
and \ref{Fdpmsqb1}.

As seen in Fig.~\ref{Fdpmsqb1his}, the majority of the mean proper motion
differences (for both dpmx and dpmy) do not exceed $\pm$2~mas/yr. 
Interestingly, at $\delta$$<$$-$30$^{\circ}$, the fraction of areas with 
relatively small mean proper motion differences is larger in both components.
Hence, the systematic differences between the proper motions in the 
2MAst and UCAC4\,HQb5 samples are typically smaller in the most southern part
of the sky, where on the other hand the random errors of 2MAst\,HQb5 are rather 
large (see Fig.~\ref{Fepm}). In this part of the sky we mention almost
symmetric distributions (with mean values of $-$0.20 and $-$0.03~mas/yr
for all mean dpmx and dpmy, respectively), whereas the upper panels of 
Fig.~\ref{Fdpmsqb1his} show clear asymmetries with more negative mean 
proper motion differences (mean values of $-$0.75 and $-$0.91~mas/yr).
This is also seen in Fig.~\ref{Fdpmsqb1}, where the yellow/red areas are
mostly located in the north or in some 
(adjacent) 
areas close to the Galactic plane.
Both yellow/red (mean differences between about $-$5 and $-$10~mas/yr)
and blue/violet (between about $+$5 and $+$10~mas/yr) areas are distributed
over many small sky patches comparable to the size of the photographic plates 
used for determining the proper motions (6$^{\circ}$ in case of PPMXL, 
5$^{\circ}$ in case of UCAC4). However, the total sky area where the
mean differences (dpmx or dpmy) exceed $\pm$5~mas/yr, is relatively small 
(only about 7\% of the common 2MAst and UCAC4\,HQb5 area). 

Extremely large mean differences (exceeding $\pm$10~mas/yr
mainly in dpmy) 
were found in only thirty 1~deg$^2$ areas.
One third of them are due to small numbers of comparison stars 
(less than 20 per 1~deg$^2$ area), and these are all located in the southern sky.
South of $\delta$$=$$-$30$^{\circ}$ all extremely 
large differences are due to small numbers of comparison stars. 
The other two thirds of areas with 
extremely large differences (all with more than 100 and up to several thousands
of comparison stars per area) are located close to the edge of the excluded 
Galactic plane in the 2MAst\,HQ5b and along zonal effects and plate edges 
such as seen in Fig.~\ref{Fskypm_p10p40}.

   \begin{figure}
   \centering
   \includegraphics[width=7.6cm]{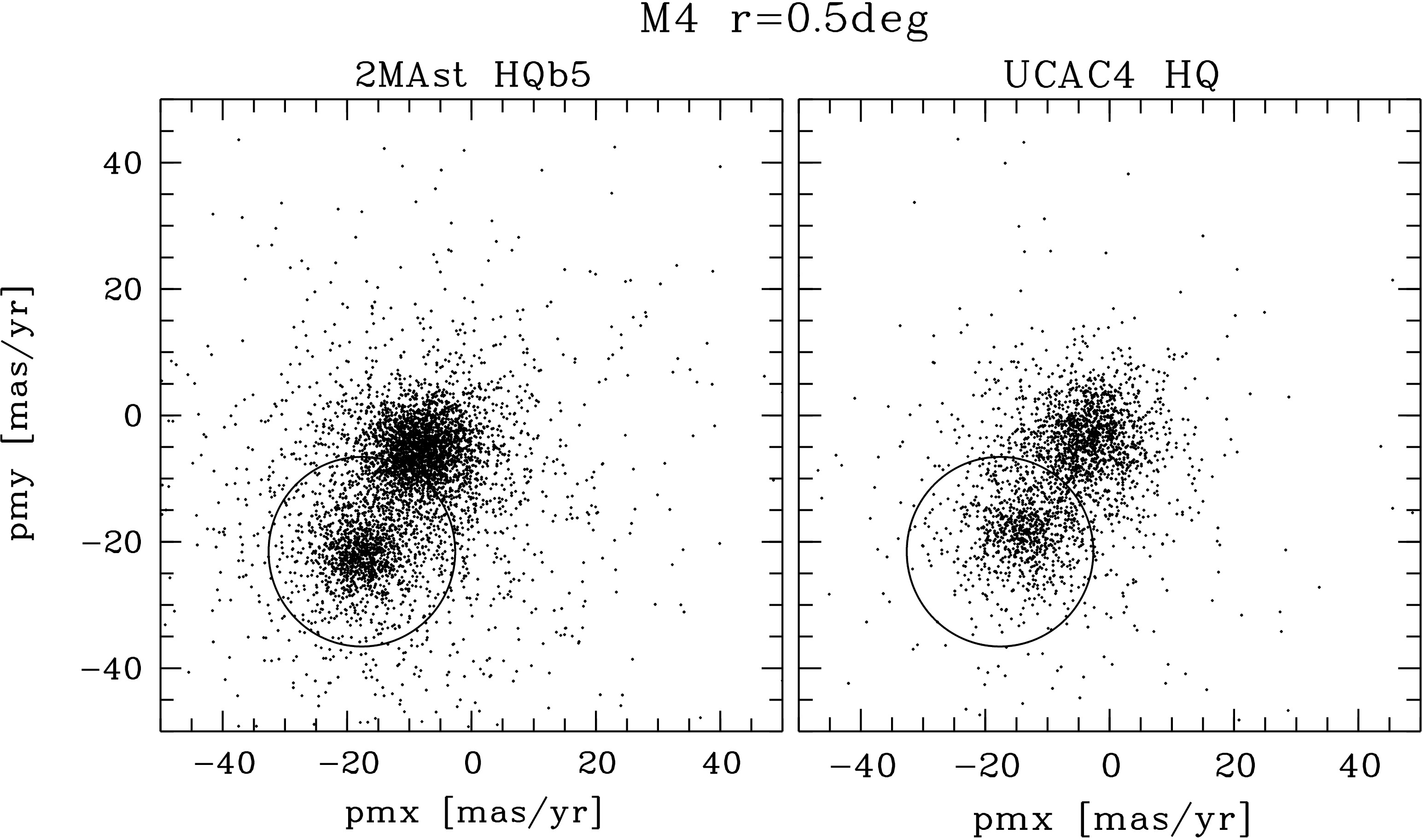}
      \caption{2MAst\,HQb5 (left) and UCAC4\,HQ (right) proper motions
               in a field centred on the globular cluster NGC\,6121 (M\,4).
               All stars within 30~arcmin from the cluster centre are
               plotted. In both panels, the circle marks the mean cluster motion as
               measured in paper II. The radius of the circle is 15~mas/yr,
               corresponding to the size of our circular proper motion bins.
               The offset of both field and cluster stars in the
               right panel illustrates the regional systematic proper motion
               differences between the catalogues discussed in 
               Sect.~\ref{SubS_comp2mastucac4}.
              }
         \label{Fm4pmcomp}
   \end{figure}

%

\section{Cluster candidate selection}\label{Sect_cand}

According to paper II (see Fig.~1 therein), the typical size (the total 
apparent radius, $r_2$) of the known clusters in the MWSC lies between 
0.06$^{\circ}$ and $\sim$0.3$^{\circ}$, with a peak at $r_2$$=$0.13$^{\circ}$. 
Only a few per cent of the 3006 
clusters with parameters determined in the MWSC have larger sizes. Among 
the new clusters found in the first MWSC extension (paper III) at relatively 
high Galactic latitudes ($|b|$$>$20$^{\circ}$), there are no clusters with
sizes $r_2$$>$0.26$^{\circ}$. For our cluster candidate selection we
binned the sky in 0.25$\times$0.25~deg$^2$ areas (again using a 
projection of $\alpha\cos{\delta},\delta$), and investigated the numbers
of stars with certain proper motions in the corresponding sky ''images''.
The idea was to identify star clusters as ''hot pixels'' in such images,
as a typical small cluster would show up as a concentration of stars
of a certain proper motion in a 0.25$\times$0.25~deg$^2$ ''pixel''
of this sky image, whereas the surrounding pixels would represent the 
background number density of field stars.

\subsection{Sky distributions of proper motion subsamples}\label{SubS_441skyimages}

As we expected no new clusters with very large proper motions
(among 2983 compact MWSC clusters from paper II
with total apparent radii $r_2$ of less than 1$^{\circ}$ only six 
have proper motions exceeding 20~mas/yr and reaching a maximum of 
about 40~mas/yr), 
we studied only 2MAst\,HQb5 stars with proper motions not exceeding 
about $\pm$50~mas/yr. 
We selected 21$\times$21 proper motion subsamples
from 2MAst\,HQb5 by shifting the central proper motion in steps of 5~mas/yr
from $-$50 to $+$50~mas/yr in each component and using a circular 
area in proper motion space with a radius of 15~mas/yr. Therefore, each 
of these 441 proper motion subsamples was strongly overlapping with 
neighbouring ones. 
We also selected the corresponding 441 proper motion subsamples from 
UCAC4\,HQ. 
We had 441$+$441 
representations of the sky, one for
each proper motion grid point, based on the 2MAst\,HQb5 and UCAC4\,HQ data,
respectively.
The relatively large radius of 15~mas/yr around the given
central proper motion of each subsample was chosen to take into account
the individual proper motion errors in 2MAst\,HQb5 (up to 8 or 12~mas/yr
in the northern and southern parts, respectively)
and UCAC4\,HQ (up to 10~mas/yr) as well as the typically small
systematic differences of a few mas/yr
between the two catalogues (see Sect.~\ref{SubS_comp2mastucac4}).
In Fig.~\ref{Fm4pmcomp}, we show as an example the proper motion
distributions of all 2MAst\,HQb5 and UCAC4\,HQ stars in a small circular 
region around the globular 
cluster NGC\,6121 (M\,4). 
The proper motion of this cluster is clearly distinct from the proper motion
of the field stars. The circle, with a radius of 15~mas/yr corresponding to
the size of our chosen circular proper motion bins, marks the cluster
proper motion as provided in paper II in both panels of Fig.~\ref{Fm4pmcomp}.
The two distributions of cluster and field stars are clearly seen. The spread
of the cluster stars and the systematic shift (in this case of the order 
of 5~mas/yr) of the UCAC4\,HQ proper motions (of both cluster and field stars) 
justify our choice of 15~mas/yr for the radius of the circular proper motion 
bins.
The Fig.~\ref{Fm4pmcomp} illustrates the regional systematic proper motion 
differences between the catalogues.

By comparing the distribution of corresponding subsamples from both
2MAst\,HQb5 and UCAC4\,HQ over the sky, and by searching for small
sky areas (pixels) with enhanced numbers of stars of a given proper motion
in both catalogues, we aimed at distinguishing real clusters from
artefacts. We expected such artefacts to occur in each catalogue
because of possible 
larger systematic proper motion errors (exceeding $\pm$5~mas/yr)
over small sky areas (cf. Fig.~\ref{Fdpmsqb1})
representing residual distorsions of the photographic plates
used for determining the proper motions in that catalogue.
With the completely different sets of photographic plates
used for PPMXL and UCAC4, the probability of finding such artefacts
in both catalogues in the same sky pixel and with the same proper motion
is extremely low.

   \begin{figure*}
   \sidecaption
   \includegraphics[width=12.9cm]{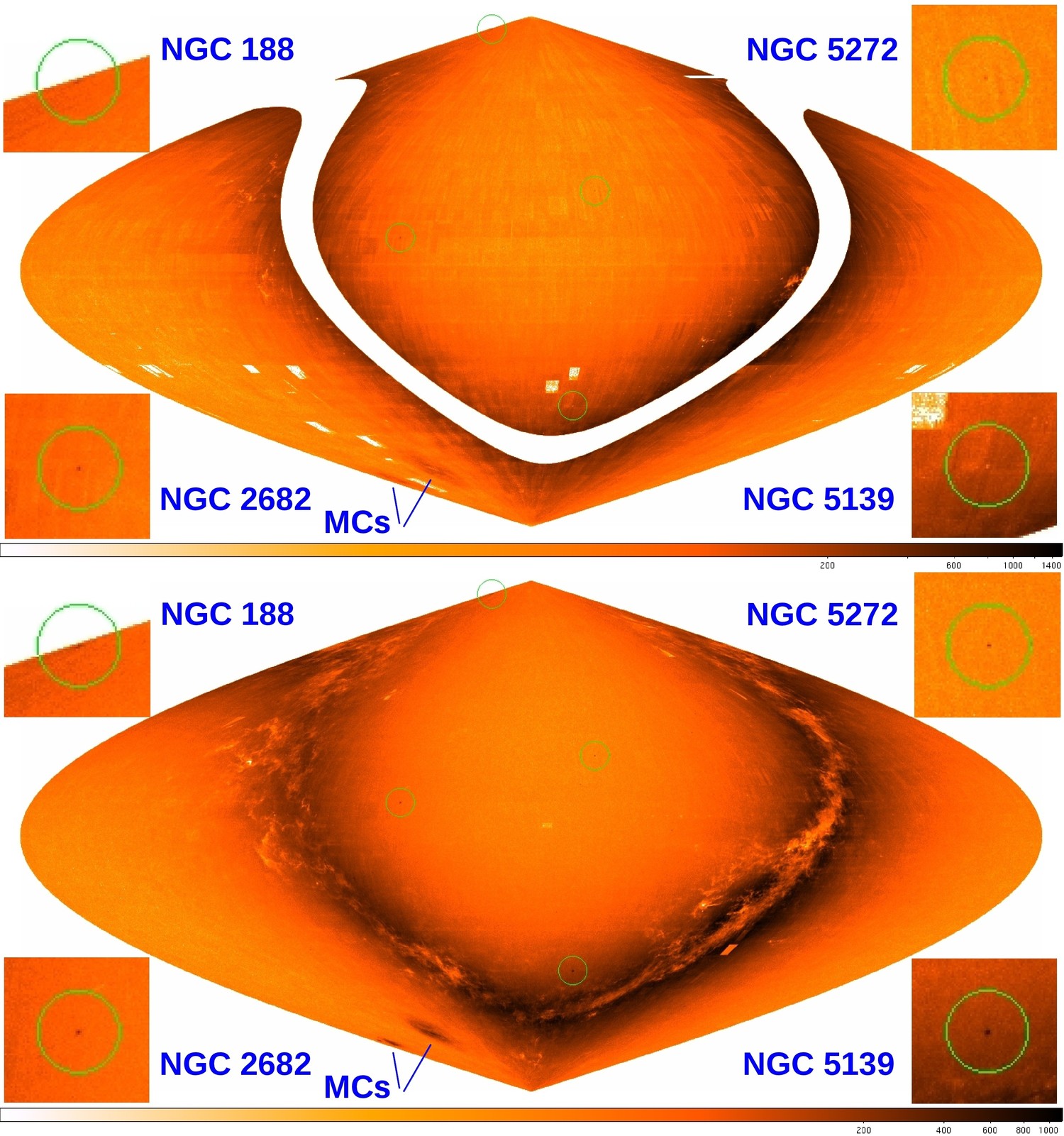}
      \caption{Density of stars with proper motions centred around
               ($\mu_{\alpha}\cos{\delta},\mu_{\delta}$)$=$($-$5,$-$5)~mas/yr
                with a radius of 15~mas/yr
               \textbf{Top:} in 2MAst\,HQb5,
               \textbf{Bottom:} in UCAC4\,HQ. Sky pixels are
               0.25$\times$0.25~deg$^2$ in a
               projection of $\alpha\cos{\delta},\delta$ (north is up, $\alpha$
               increases from left to right). The maximum numbers of objects
               per pixel are slightly more than 1400 in 2MAst\,HQb5 and 1000
               in UCAC4\,HQ, respectively.
               Open circles mark the positions of the open cluster NGC\,188,
               the globular cluster NGC\,5272 (M\,3), the open cluster
               NGC\,2682 (M\,67), and the globular cluster
               NGC\,5139 ($\omega$ Cen)
               (from north to south), also shown in the zoomed images at
               the corners. 
               Also marked are the Magellanic Clouds.
              }
         \label{Fskypm_m05m05}
   \end{figure*}

For all 441$+$441 proper motion subsamples from 2MAst\,HQb5 and UCAC4\,HQ
we created the corresponding sky images with 0.25$\times$0.25~deg$^2$
pixel sizes, where the number of objects falling into a sky pixel was 
transformed to its intensity. These 882 images were created 
by using the European Southern Observatory (ESO) 
Munich Image Data Analysis System (MIDAS) tool convert/tab.
The output in the
format of the Flexible Image Transport System (FITS) formed the 
basis for our cluster candidate search. In Fig.~\ref{Fskypm_m05m05} we
show as an example two sky images for subsamples with the same relatively 
small proper motions, as they appear based on 2MAst\,HQb5 and UCAC4\,HQ data, 
respectively. For comparison, 
Figs.~\ref{Fskypm_p10p40} and \ref{Fskypm_p20m45} show the corresponding
images for two subsamples with relatively large proper motions.
The Figs.~\ref{Fskypm_m05m05}, \ref{Fskypm_p10p40}, 
and \ref{Fskypm_p20m45} are discussed in more detail in the next section.

\subsection{Visual detection of clusters}\label{SubS_visidetect}

\begin{figure*}
   \sidecaption
   \includegraphics[width=12.9cm]{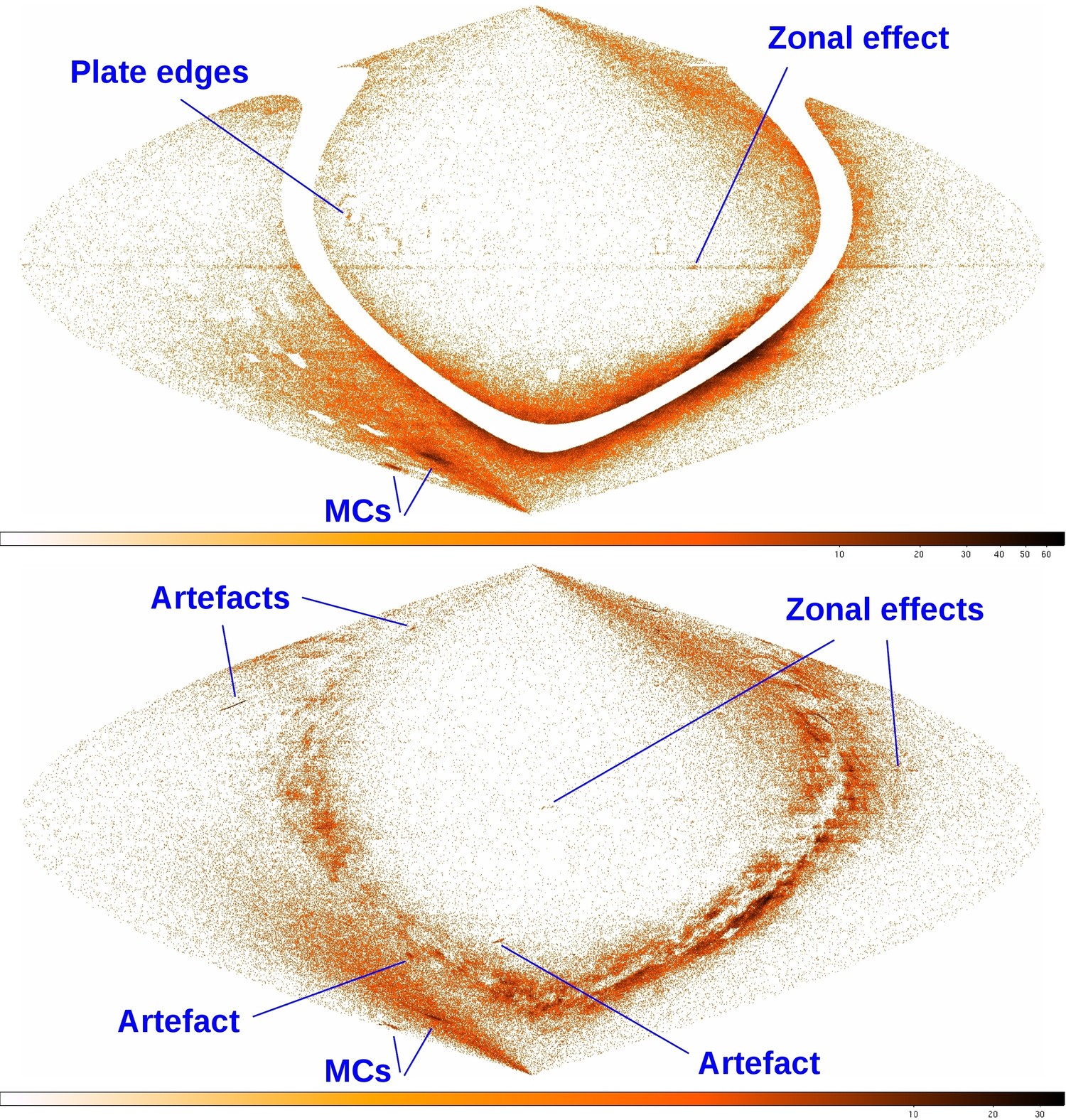}
      \caption{As Fig.~\ref{Fskypm_m05m05} but for
               stars with relatively large proper motions centred around
               ($\mu_{\alpha}\cos{\delta},\mu_{\delta}$)$=$($+$10,$+$40)~mas/yr
                with a radius of 15~mas/yr
               \textbf{Top:} in 2MAst\,HQb5 (maximum number of objects
               per 0.25$\times$0.25~deg$^2$ pixel is about 60),
               \textbf{Bottom:} in UCAC4\,HQ (maximum of about 30).
               The Magellanic Clouds (MCs) are marked.
               Zonal effects at the
               edges of photographic plates used for the proper motion
               determination are seen in both plots. Some UCAC4
               artefacts are marked.
              }
         \label{Fskypm_p10p40}
   \end{figure*}

\begin{figure*}
   \sidecaption
   \includegraphics[width=12.9cm]{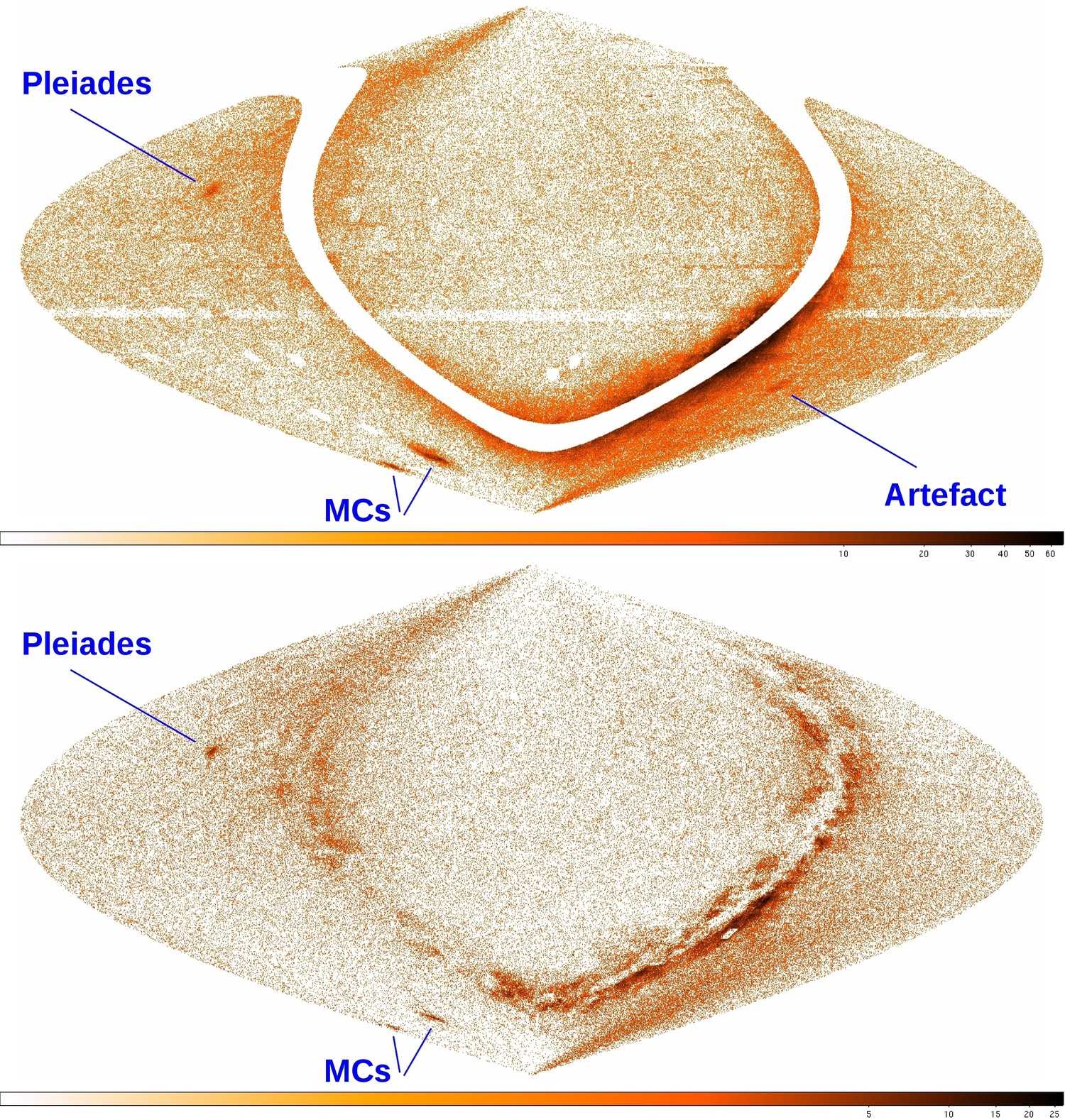}
      \caption{As Fig.~\ref{Fskypm_m05m05} but for
               stars with with relatively large proper motions centred around
               ($\mu_{\alpha}\cos{\delta},\mu_{\delta}$)$=$($+$20,$-$45)~mas/yr
                with a radius of 15~mas/yr
               \textbf{Top:} in 2MAst\,HQb5 (maximum number of objects
               per 0.25$\times$0.25~deg$^2$ pixel is about 60),
               \textbf{Bottom:} in UCAC4\,HQ (maximum of about 25).
               The Magellanic Clouds (MCs) are marked.
               The Pleiades can easily be seen
               in both plots. A southern 2MAst\,HQb5 artefact,
               similar in appearance
               to the Pleiades, is also marked. 
              }
         \label{Fskypm_p20m45}
   \end{figure*}

For the visual inspection of the large numbers of FITS images created
(see Sect.~\ref{SubS_441skyimages}), we used
the SAOImage DS9 (Joye \& Mandel~\cite{joye03}) tool. We found this in
particular very useful, as it not only allowed us to blink the corresponding
2MAst\,HQb5 and UCAC4\,HQ sky images with different colour maps, 
zoom factors, etc., but 
also to effectively watch short movies of a series of images with similar 
(overlapping) proper motions. Sky pixels 
popped up with more or less constant high intensity in
such movies, when they
contained 
known clusters and new cluster candidates.
The inverted colour map ''heat'' with a logarithmic scale chosen 
for Figs.~\ref{Fskypm_m05m05}, \ref{Fskypm_p10p40} and \ref{Fskypm_p20m45}
shows the full dynamic range of these images, which is very different for
the large numbers of objects with small proper motions compared to much
less objects with large proper motions, where many sky 
pixels are empty.

Window-like 
patterns of the photographic plates used for determining
the proper motions can be seen in the sky images of both catalogues. The
different observational history (various epoch differences) over the sky
resulted in different levels of proper motion errors. Therefore, we found
even stronger plate patterns in sky images representing objects with small
proper motions like in Fig.~\ref{Fskypm_m05m05}, when we tried to use more 
stringent criteria for the proper motion errors, e.g. UCAC4 errors of 
less than 7~mas/yr instead of our chosen limit of 10~mas/yr.
The choice of the two different error limits (Sect.~\ref{SubS_ppmxl})
for 2MAst\,HQb5 objects north and south of $\delta$$=$$-$30$^{\circ}$
led to a smooth transition in the number density of objects accross that
border. Declination zones and plate borders with density enhancements
of stars with large proper motions can also be seen in both catalogues
(Figs.~\ref{Fskypm_p10p40} and \ref{Fskypm_p20m45}). These patterns are
not a consequence of our error cuts but hint at systematically wrong 
proper motions because of improper treatment of some of the photographic 
plates involved in 
each of the
catalogues. In cases, where whole
photographic plates, their edges, or corners are affected, the
resulting apparently large proper motions mimic large clusters or streams
as the artefacts marked in Figs.~\ref{Fskypm_p10p40} and \ref{Fskypm_p20m45}.

While we excluded the Galactic plane in the 2MAst\,HQb5 sample, we
did not exclude the very crowded fields of the Magellanic Clouds. 
Interestingly, they appear 
with a much weaker signal
in the upper part of 
Fig.~\ref{Fskypm_m05m05} than in the lower part. We also see in the zoomed
subimages centred on the largest globular cluster in the sky, NGC\,5139 
($\omega$ Cen), that there is a minimum in the 2MAst\,HQb5 number density,
whereas the UCAC4\,HQ image shows a clear peak. This means that the Schmidt 
plates used for the PPMXL (and 2MAst) did not allow to determine reliable
proper motions (which are small for the Magellanic Clouds and for
NGC\,5139) for the majority of objects in very crowded regions. On the other
hand, we see the Magellanic Clouds as clear density enhancements in the
sky images representing objects with large proper motions
(Figs.~\ref{Fskypm_p10p40} and \ref{Fskypm_p20m45}) from both catalogues
(although they are stronger in case of 2MAst\,HQb5). We also 
noticed
a (weaker) peak at the position of NGC\,5139 in other UCAC4\,HQ
sky images (not shown here) over a relatively wide range of proper motions.
This effect may be because of crowding 
problems in dense regions leading 
to concentrations of objects with any kind of proper motions. 
These problems 
affect the 2MAst\,HQb5 proper motion subsamples to a higher degree 
than those from 
the UCAC4\,HQ.

The 
chosen 
mean proper motion of the subsamples 
shown in Fig.~\ref{Fskypm_m05m05} is according to the results of paper II
very similar to the known mean proper motions of three of the marked clusters 
NGC\,5139 ($\mu_{\alpha}\cos{\delta},\mu_{\delta}$$=$$-$6.0,$-$5.0~mas/yr),
NGC\,2682 ($-$7.3,$-$5.9~mas/yr), NGC\,5272 ($-$5.6,$-$6.1~mas/yr), but
also close to that of the fourth, NGC\,188 ($+$0.2,$-$1.4~mas/yr). Besides of 
the already mentioned strong crowding problems in the 2MAst\,HQb5 data
for NGC\,5139, the peaks for the other three clusters are clearly weaker
in 2MAst\,HQb5 than in UCAC4\,HQ, although the dynamic range of the UCAC4\,HQ 
image is lower. This is a representative example, as we usually observed 
a stronger signal for 
well-known rich
clusters in the UCAC4\,HQ data compared to the 
2MAst\,HQb5 data. Therefore, in our comparative search for clusters by proper 
motion and concentration in small sky areas, the higher number of objects 
in the 2MAst\,HQb5 sample was apparently compensated by the reduced crowding 
problems in the UCAC4\,HQ data.

The mean proper motion of the subsamples shown in Fig.~\ref{Fskypm_p20m45}
is approximately equal to that of the Pleiades (Melotte\,22), which show up 
clearly in both 2MAst\,HQb5 and UCAC4\,HQ data. This is an example for the few 
known extended clusters
(with total apparent radii, $r_2$, clearly exceeding 1$^{\circ}$ in paper II)
easily visible in those of our 882 sky images, where the proper motion
of the cluster overlaps with our
chosen circular proper motion bins. 
Other easily visible extended clusters (not shown) 
were Praesepe (NGC\,2632) and 
the $\alpha$\,Per cluster (Melotte\,20). On the other hand, extended
clusters like Blanco~1 or Ruprecht\,147 (also discussed as problematic 
in paper III) were difficult to detect visually (in case of Ruprecht\,147
we visually detected the core of the cluster in one sky pixel) and fell
below our thresholds with our final 
automated detection 
(Sect.~\ref{SubS_autodetect}). 
However, we did not intend to discover new extended clusters in this work, 
so our automated 
detection aimed at finding compact clusters.

By visual inspection, we found about 120 compact 
(concentrated in one or a few neighbouring pixels) cluster 
candidates, half of which turned out to be known open and globular clusters.
These visually identified known clusters and cluster candidates served
as comparison objects in the process of developing 
the automated procedure
for the compact cluster detection described in the next section.

   \begin{figure*}
   \centering
   \includegraphics[width=15.0cm]{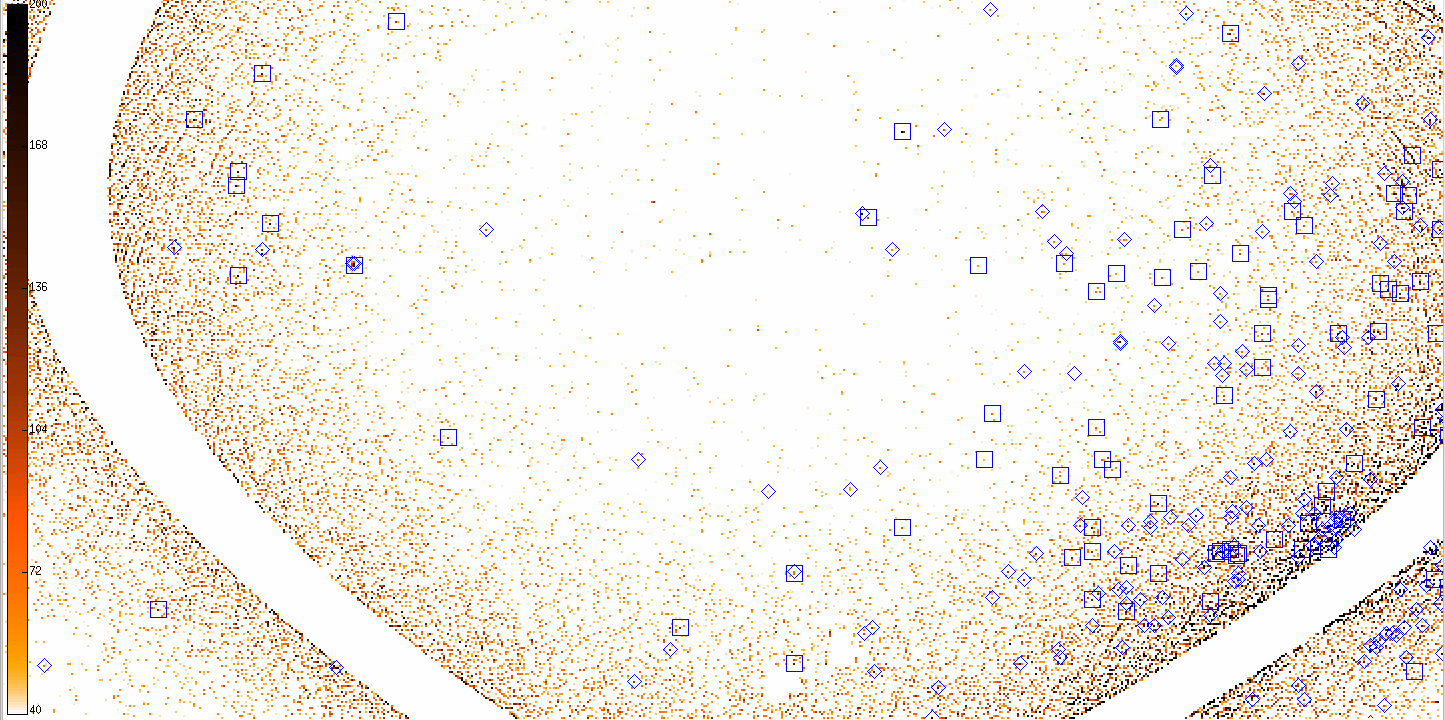}
   \includegraphics[width=15.0cm]{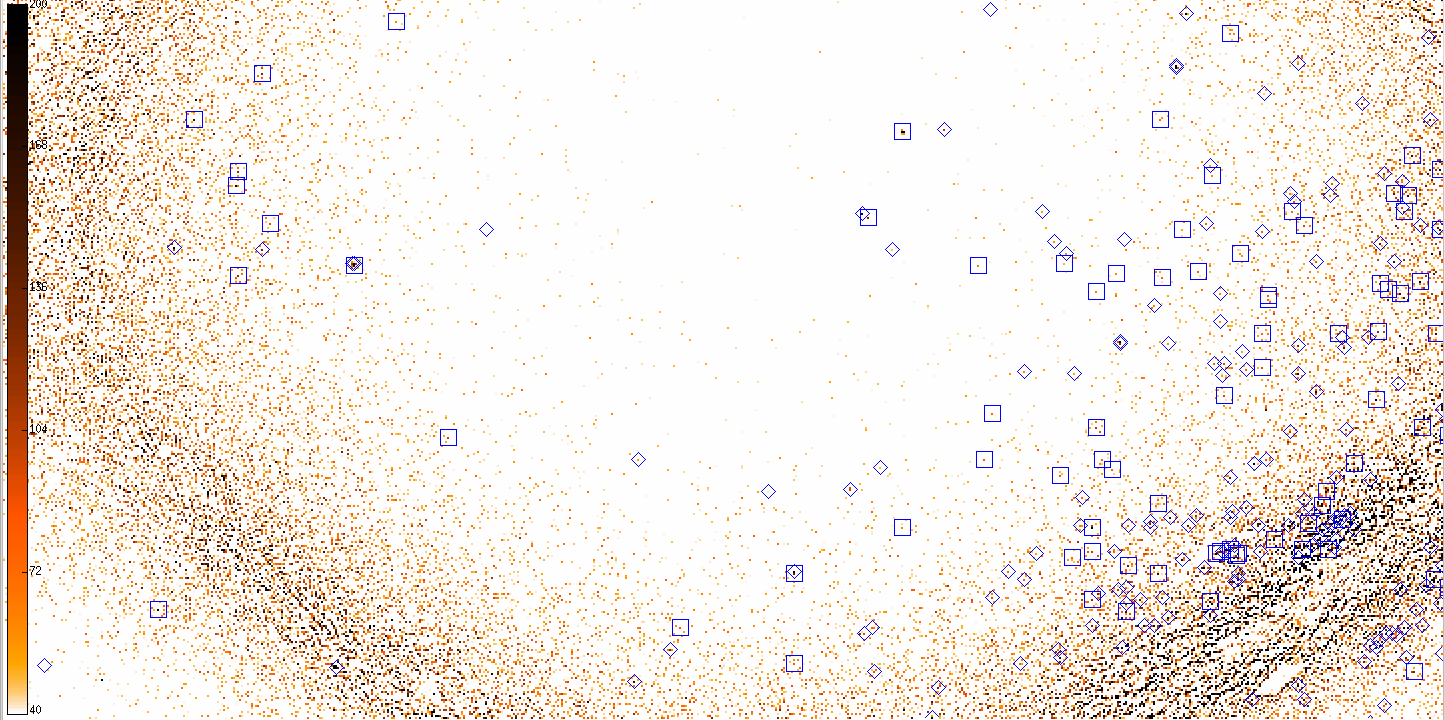}
   \includegraphics[width=15.0cm]{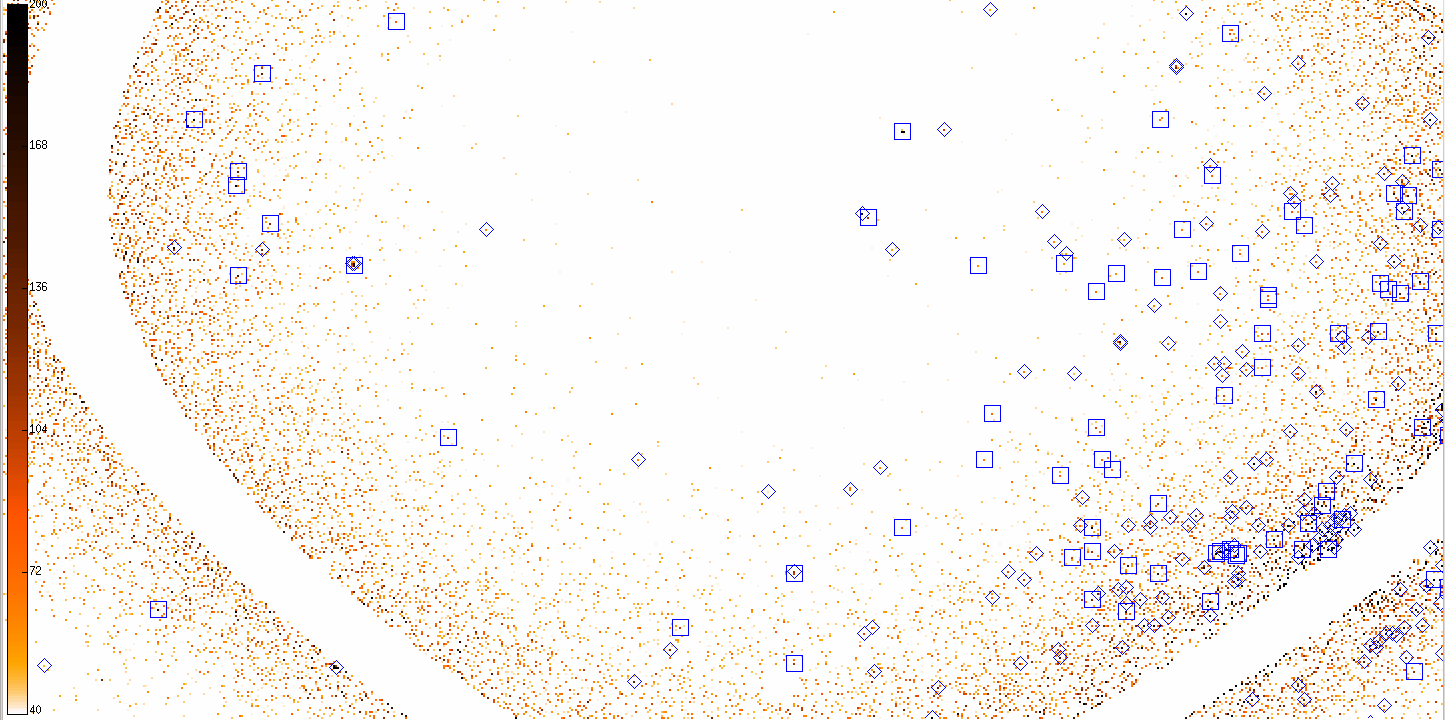}
      \caption{Zoomed view on the central part of the sky
               (projected as in Fig.~\ref{Fskypm_m05m05}) 
               after step 1 described in Sect.~\ref{SubS_autodetect}.
               The intensities are equivalent to the sum of
               four original sky images after background subtraction.
               In this example, the central proper motion grid point is at
               ($\mu_{\alpha}\cos{\delta},\mu_{\delta}$)$=$($-$2.5,$-$2.5)~mas/yr,
               similar to the proper motion bin shown in Fig.~\ref{Fskypm_m05m05}.
               The shown filtered images represent: Top: set A (sum of four 
               2MAst\,HQb5 images),
               Middle: set B (UCAC4\,HQ sum), and Bottom: set AB (the 
               square root of the product of the two sums, respectively).
               Open squares mark the cluster candidates found after step 4
               of Sect.~\ref{SubS_autodetect} with the given central proper motion.
               Open lozenges mark candidates with similar (overlapping) proper
               motions.
              }
         \label{Ffiltsky}
   \end{figure*}

\subsection{Automated detection of clusters}\label{SubS_autodetect}

During visual inspection, known clusters 
and candidates showed up as bright sky pixels 
in several overlapping proper motion subsamples of both the 2MAst\,HQb5 
and UCAC4\,HQ image sets. Because of the large dynamic range of the sky 
images, in particular those representing small proper motions, and changing 
background (density of field stars) over the sky, the visual detection 
profited from the use of different colour maps and scalings (in SAOImage DS9). 
Our automated 
procedure 
aimed at detecting compact cluster candidates
as 'hot' sky pixels in proper motion-selected representations of the sky.
Therefore, it appeared logical to combine the original sky images, apply 
filters for background subtraction and set thresholds corresponding to a
minimum of possible cluster stars in a sky pixel.
Summing up the sky images corresponding to strongly overlapping
proper motion subsamples had the clear advantage of enhancing the signal
and reducing the noise in our cluster search with still high enough resolution
in the proper motion space.
We performed the following steps in this 
automated detection process:

\begin{enumerate}
  \item
Using MIDAS tools for operations on images, we created 
three sets 
of images: one from the original 441 2MAst\,HQb5 
images (set A), a second from 441 UCAC4\,HQ images (set B), and a third from 
combining the 441$+$441 images (set AB). 
For each image of sets A and B, 
we combined four of the original sky images corresponding to proper motion 
subsamples overlapping in both directions. 
For each image of set AB,
we combined eight of them. 
The new central proper motion grid points were separated by 10~mas/yr.
The pixel intensities 
in the sky images
of sets A and B were computed as the sum of four 
background-corrected original intensities,
where the background around each pixel was determined as the 
median intensity of 
the eight
surrounding pixels. Negative
sums were set to zero. 
To avoid artefacts appearing in only one of the catalogues,
the more important set of combined images (image set AB) was created with 
intensities
computed as the square root of the product of 
these two sums from 2MAst\,HQb5 and UCAC4\,HQ, respectively.
  \item Subsequently, using the MIDAS tool find/pixel, we created three 
tables (tables A, B, and AB) containing all
intensities 
for all sky pixel locations and for all images 
(corresponding to the new proper motion grid points)
of the three image sets, 
respectively.
We considered table AB as our main source for
the cluster search by proper motions, as it contained the combined
proper motion information from both catalogues. 
High intensities in only one of the proper motion catalogues 
(artefacts) that were not above the background in the other led to
zero intensities in table AB. Note that the maximum intensity for 
a given sky pixel in table AB
was the result of a large number of stars in the same proper motion bin
for both 2MAst\,HQb5 and UCAC4\,HQ. On the other hand, if a given
sky pixel intensity was at maximum in both tables A and B, this did
not necessarily mean that these maxima corresponded to the same proper motion
bin. Although, these proper motion bins were typically overlapping.
  \item As the values in the intensity tables equal
sums of four original representations of the sky corrected for the
background, we applied a threshold of 49 for table AB,
corresponding to more than 12 possible cluster stars
of a given proper motion. 
For the independent tables A and B, we used a slightly 
lower threshold of 40, 
corresponding to a 
minimum of 10 possible cluster stars
above the local background numbers in the original images. By doing so, 
we allowed for small variations taking
into account
the different magnitude-dependent object numbers in our proper motion samples 
(Fig.~\ref{FKs}) as well as the different performance of the two proper
motion catalogues in crowded regions mentioned in Sect.~\ref{SubS_visidetect}.
Only sky pixels with 
intensities 
above the corresponding thresholds in all 
three tables were further considered as potential locations of star clusters.
  \item Despite the correction for the background performed 
in step 1,
the number of 
candidates 
was strongly increasing towards low
Galactic latitudes. Therefore we applied an additional latitude-dependent
threshold of 45$/|\sin{b}|$ to the
intensities 
in table AB, preserving the equivalent 
minimum object number of about 12 only at high Galactic latitudes
($|b|>$60$^{\circ}$). Our requested minimum object numbers per sky pixel 
increased to e.g. about 18 at $|b|=$40$^{\circ}$, 33 at $|b|=$20$^{\circ}$,
and raised to about 65 at $|b|=$10$^{\circ}$, and the maximum of about
130 at $|b|=$5$^{\circ}$. The final number of 
cluster candidates was 692. All thresholds
were adjusted so that the success rate of recovering
the visually found clusters and candidates 
was at maximum.
\end{enumerate}

In Fig.~\ref{Ffiltsky} we show examples of filtered images 
of sets A, B, and AB (after step 1) 
with a 2$\times$ zoom on the central part of 
the sky projection used in Figs.~\ref{Fskypm_m05m05}, \ref{Fskypm_p10p40},
and \ref{Fskypm_p20m45}. 
The same intensity cuts were applied to all three images, as the 
intensities represent the background-corrected sums of four original images. 
The intensity of bright sky pixels can be considered as the fourfold number 
of member stars of a cluster candidate. There are many bright sky 
pixels appearing only in one of the upper two images, representing the
2MAst\,HQb5 and UCAC4\,HQ data independently. The lower image shows the
result of combining both data sets. It is still possible that
a very large number in only one of the catalogues multiplied by a small
number above zero in the other leads to a cluster candidate in the
combined data set AB. However, applying all three thresholds to sets
A, B, and AB, as described in step 3, we effectively excluded such cases. 
In addition to the marked candidates in Fig.~\ref{Ffiltsky} with
proper motions equal or similar to the given proper motion bin, one
can see even more relatively bright sky pixels, especially near 
the Galactic plane, at the same location in all three images. With our
last Galactic latitude-dependent criterion in step 4, we excluded such
cases, possibly resulting from noise in dense regions of the sky.

   \begin{figure*}
   \centering
   \includegraphics[width=15.3cm]{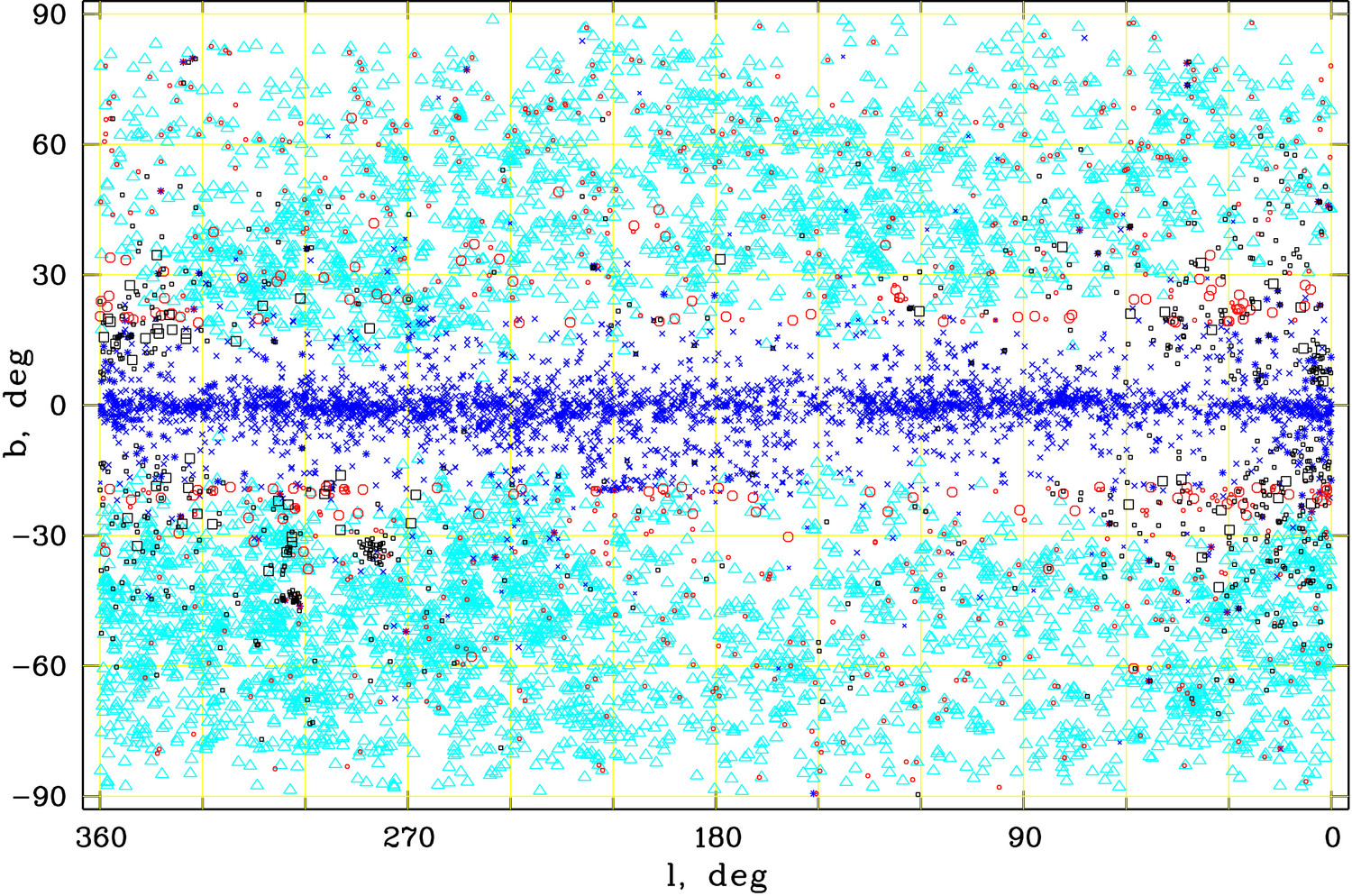}
      \caption{Distribution of all MWSC clusters and candidates
               in Galactic coordinates:
               blue crosses - MWSC open clusters (large) and candidates (small)
                              from paper II,
               blue asterisks - MWSC globular clusters (large) and
                                candidates (small) from paper II,
               red open circles - clusters (large) and candidates (small)
                                  from paper III,
               black open squares - newly discovered clusters (large) and
                                    all other candidates (small)
                                    from this paper,
               cyan open triangles - clusters of galaxies
                                     (Abell et al.~\cite{abell89}).
              }
         \label{Fcl_and_cand_lb}
   \end{figure*}

The proper motions of the 692 candidates 
were mostly small. 
Only for about 45\% of
the cluster candidates, one of the proper motion components was $\pm$12.5
or larger, reaching a maximum of $\pm$22.5~mas/yr in very few cases.
These proper motions were considered only as rough initial estimates, 
which had still to be improved by the pipeline (Sect.~\ref{Sect_new}).
The initial size of all our cluster candidates corresponded 
to the pixel size (0.25~arcsec),
but there were also adjacent hot pixels in the sky images, representing
larger known clusters and cluster candidates.
The largest known cluster detected 
with the automated procedure 
was NGC\,2682 with a
total apparent radius of $r_2$$=$1.03$^{\circ}$ (according to paper II).
The pixel coordinates provided only a relatively uncertain initial position
of the candidates for their further analysis (Sect.~\ref{Sect_new}).
The distribution of all 692 candidates, including the later confirmed
63 new clusters (Sect.~\ref{Sect_new}), 
can be seen 
in Fig.~\ref{Fcl_and_cand_lb}. Most of our candidates
were concentrated in a wide region around the Galactic centre,
although our final threshold was dependent only
on Galactic latitude, not longitude.

%
\begin{table}[b]
\caption{Identification of our cluster candidates} 
\label{Tidcand}      
\centering                          
\begin{tabular}{rlr}        
\hline\hline                 
Number       & identified                         & Number        \\
of           & with                               & of            \\
candidates   & object type                        & objects       \\
\hline
155\,\,\,          & known clusters of galaxies& 150  \\
 67\,\,\,          & known globular clusters            &           46  \\
 56\,\,\,          & known open clusters (paper II)     &           47  \\
 21\,\,\,          & known open clusters (paper III)    &           21  \\
  6\,\,\,          & unconfirmed candidates (paper III) &            6  \\
67\,\,\,  & new open clusters (this paper)     &  63  \\
 33\tablefootmark{a}          & Large Magellanic Cloud             &            1  \\
 17\tablefootmark{a}          & Small Magellanic Cloud             &            1  \\
270\,\,\,          & not confirmed             &          270  \\
\hline                                   
\end{tabular}
\tablefoot{
\tablefoottext{a}{These cluster candidates overlap 
with the very crowded regions of the Magellanic Clouds,
where the 2MAst\,HQ proper motions are unreliable.
}
}
\end{table}

%

\section{Known clusters among the candidates}\label{Sect_known}

The efficiency of our 
automated 
cluster candidate selection was proved by 
the fact that 355 (51\%) of the 692 cluster candidates turned out to overlap 
with previously known Galactic star clusters (or candidates), the Magellanic
Clouds, or known 
clusters of galaxies 
(Table~\ref{Tidcand}). Several candidates 
overlapped with the same known clusters that occupy relatively 
large areas on the sky. Interestingly, 
21 of our candidates were also found and confirmed as open clusters in
paper III, whereas another six common candidates of both studies were
not confirmed by the MWSC pipeline. 
We discovered 63 new open clusters (Sect.~\ref{Sect_new}), 
corresponding to 67 of our candidates, partly overlapping with each other. 
Among our 692 cluster candidates, there were 270 (39\%) that
could neither be identified with known
objects nor confirmed by the MWSC pipeline.

There are higher numbers of 
known globular and especially open clusters 
among our 692 candidates 
compared to the numbers of known Milky Way clusters among the 782
candidates of paper III. This is probably caused by extending our search to
lower Galactic latitudes. The smaller contamination 
of our survey by
clusters of galaxies is probably
achieved because of the same reason (clusters of galaxies become important 
at Galactic latitudes exceeding $\pm$15$^\circ$). The reduced contamination 
may also be related 
to our search by proper motions, although most of our 
candidates were found within the smallest proper motion 
bins. If we compare 
the numbers of all known globular (109) and open clusters (863)
in our search area ($|b|$$>$5$^\circ$) with the corresponding numbers of
detected objects in Table~\ref{Tidcand}, we find that our search was much 
more sensitive to globular clusters than to open clusters. The 46 detected
globulars correspond to a success rate of 42\%. After excluding all open
clusters more extended than our largest detected open cluster, NGC\,2682,
and all associations, moving groups, embedded clusters and cluster remnants,
there are still 575 objects in the MWSC catalogue of paper II and 128 objects
from paper III. 
Our success rate 
for open clusters is then only about 10\%. 
The success rates for globular and open clusters
are smaller than in paper III, probably because of the low quality of 
the currently available proper motions.

%

\section{New clusters confirmed by the pipeline}\label{Sect_new}

The reduction pipeline combining the stellar positions and proper motions
from the PPMXL catalogue with photometry from the 2MASS catalogue
for cluster membership and parameter determination was described in paper I.
We have not used the proper motions from UCAC4 in the pipeline, to be
consistent with our MWSC survey.
The main parameters (equatorial coordinates RA and Dec, Galactic coordinates
$l$ and $b$, the total apparent cluster radii $r_2$ and the numbers $n_2$
of 1-$\sigma$ members 
(highly probable cluster members)
within $r_2$, the proper motion components pmx and
pmy, the distances $d$, and the ages $\log{t}$) of the 63 new clusters are 
listed in Table~\ref{Tmainpar}. The full tables with all determined 
cluster parameters in the same format as provided in paper I, II, and III 
are again available 
in electronic form only at the CDS. There one can also find
the atlas pages with the standard MWSC diagrams and the membership lists 
for each of the new MWSC clusters.
To distinguish the clusters and candidates of our proper motion 
search from previous MWSC entries, we assigned our objects with MWSC numbers 
in the range 4000...4999.

   \begin{figure}
   \centering
   \includegraphics[width=8.8cm]{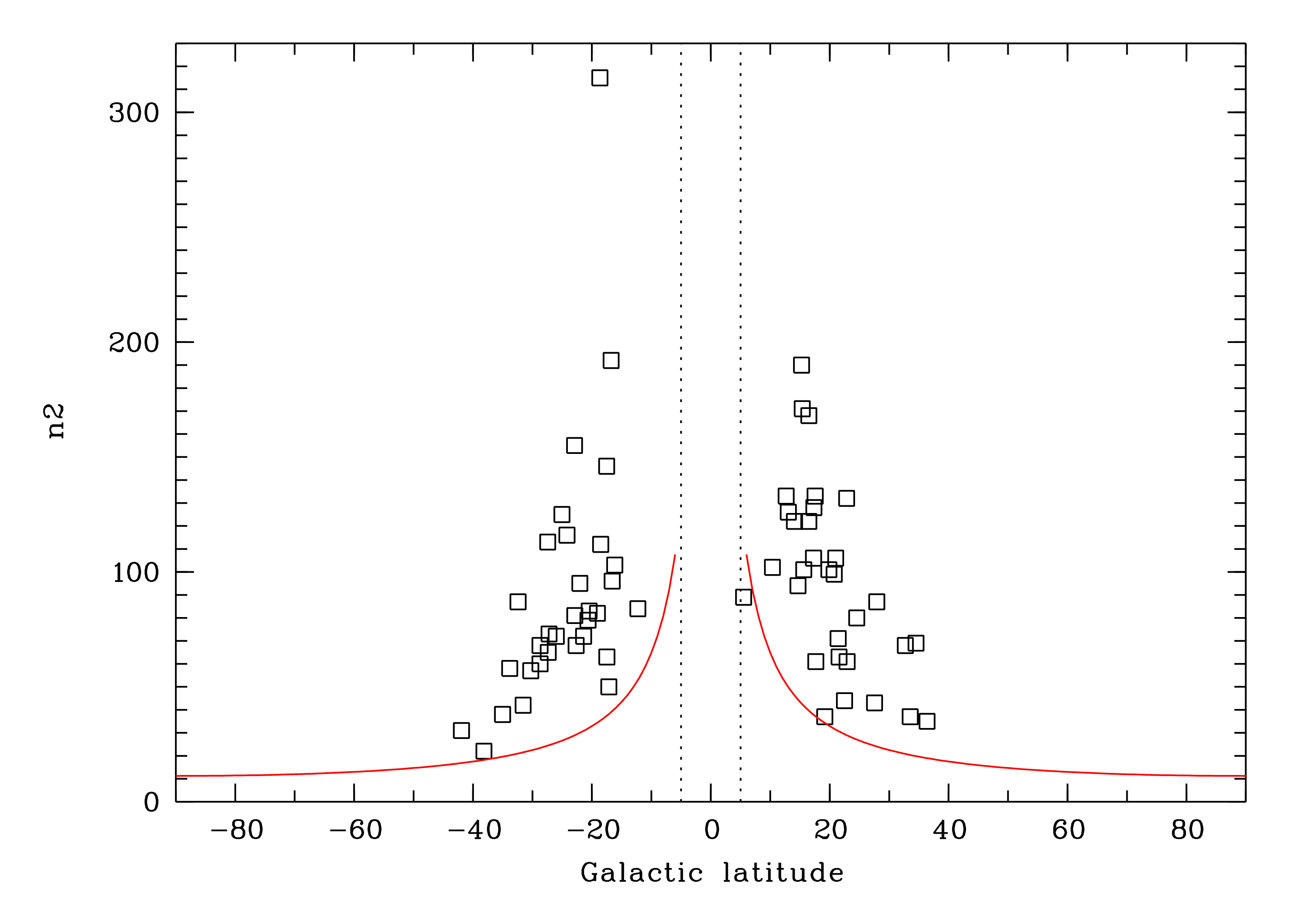}
      \caption{Number of highly probable members ($n_2$) for 63 new
               clusters confirmed by the MWSC pipeline as a function
               of Galactic latitude. The excluded Galactic plane zone
               is indicated by dotted lines. The red solid lines
               correspond to the threshold set for our cluster
               candidates
               in step 4 of Sect.~\ref{SubS_autodetect}.
              }
         \label{Fbn2}
   \end{figure}

   \begin{figure*}
   \sidecaption
   \includegraphics[width=12.9cm]{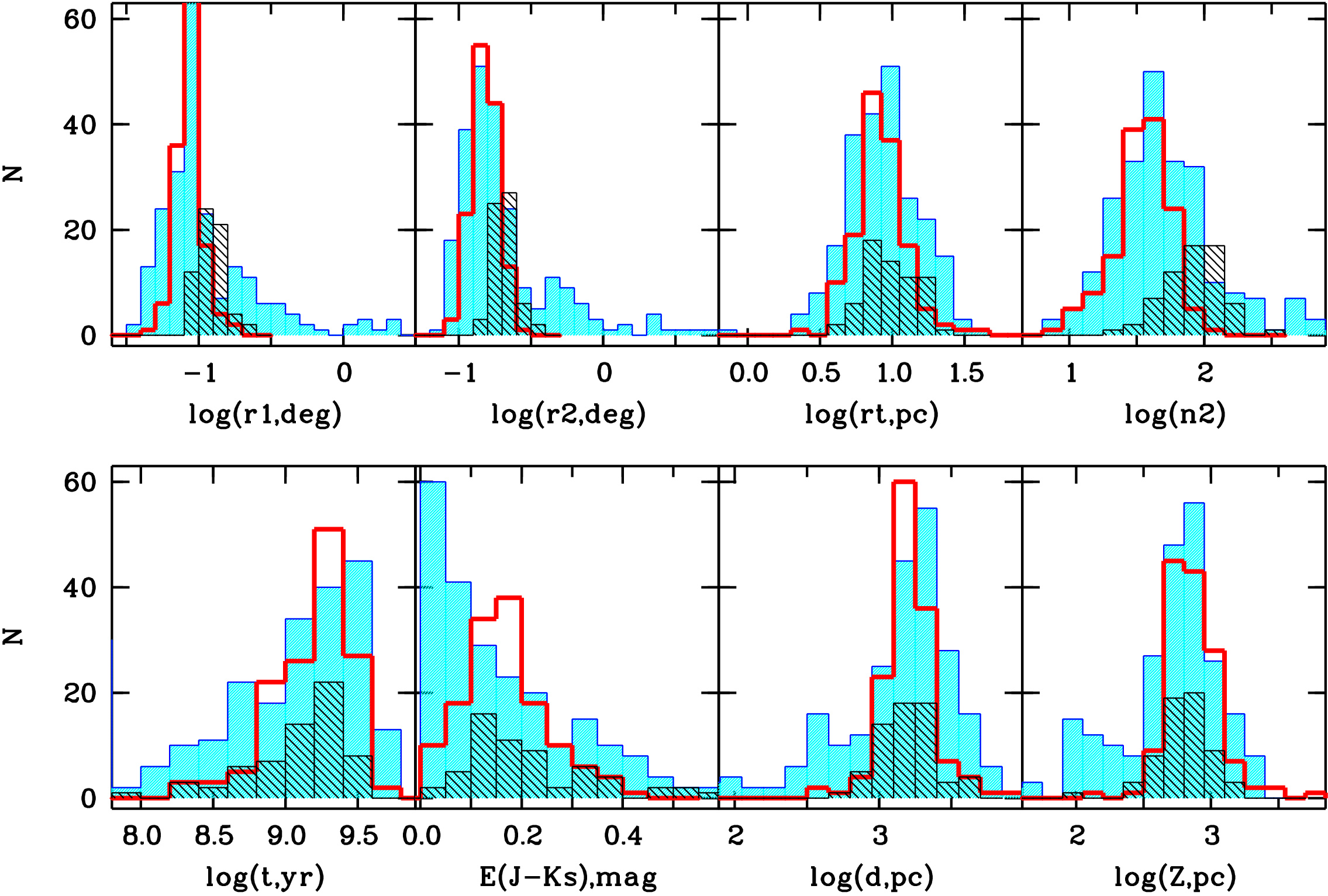}
      \caption{Distribution of determined cluster parameters ($r_1$ - angular
               radius of the central part, $r_2$ - total apparent radius,
               $r_t$ - tidal radius, $n_2$ - number of 1-$\sigma$ members
               within $r_2$, $t$ - age, $E(J-K_s)$ - colour excess,
               $d$ - distance, $Z$ - height above the Galactic plane):
               blue (filled histogram) - 231 MWSC open clusters
               with $|b|>$15$^{\circ}$ from paper II,
               red (thick line) - 139 new open clusters from paper III,
               black (striped) - 63 newly confirmed open clusters (this paper).
              }
         \label{Faiparihis}
   \end{figure*}

Table~\ref{Tmainpar} and Fig.~\ref{Fbn2} show that only one of the 63
new open clusters lies at $5^{\circ}<|b|<10^{\circ}$ and the maximum
of $|b|$ 
is at about 42$^{\circ}$. Only 26 (17 globulars
and 9 open clusters) of all 3006 MWSC clusters 
(paper II)
are located 
at $|b|>40^{\circ}$. We added one new open cluster in that part of the 
sky, whereas paper III provided seven discoveries, respectively.
As can be seen in Fig.~\ref{Faiparihis}, the distributions of
several parameters of our new 63 clusters, namely the tidal radii, ages,
distances, and heights above the Galactic plane, fit well within the
corresponding main bodies of the MWSC distributions for clusters 
at moderately high Galactic latitudes ($|b|>$15$^{\circ}$). The distribution 
for the determined 
reddening parameters $E(J-K_s)$ 
shows the same shift with respect to
the MWSC distribution as the corresponding distribution of the new
higher-latitude ($|b|>$20$^{\circ}$) clusters from paper III. 
This can be in part explained by the higher average interstellar
extinction towards the Galactic centre, where we have found the bulk of 
our candidates. On the other hand, this effect may be just due to the
fact that most of the less-reddened clusters were already discovered
in the past.

We classified about 20\% of the new clusters as 
sparse clusters or clusters with poor radial density profiles (RDP)
(Table~\ref{Tmainpar}). We typed one of these clusters (MWSC\_4131)
as a cluster remnant and considered it as a cluster candidate.
For comparison, 584 of the 3006 MWSC clusters (19\%)
of paper II were noted to be sparse or very sparse clusters,
and 208 of the 3006 MWSC clusters (7\%) as having a poor RDP
(some have both notes). 168 of the 3006 MWSC clusters (6\%)
were typed as remnant clusters and cluster candidates. Thus
our rate of ''problematic'' clusters is comparable or 
slightly smaller than that of paper II.

Although our 
automated detection 
of cluster candidates
aimed at finding relatively compact clusters, 
we see a trend
in Fig.~\ref{Faiparihis} to higher values
for our determined radii $r_1$ and $r_2$, as well as for the numbers $n_2$
of 1-$\sigma$ members within the total apparent cluster radius, 
compared to the corresponding typical values of the clusters
outside the immediate Galactic plane regions in papers II and III.
The numbers 
of  1-$\sigma$ members
found by the MWSC pipeline are similar to the numbers of stars
estimated for the corresponding candidates. As can be seen in Fig.~\ref{Fbn2},
these numbers lie, with only one exception, clearly above the threshold
that was dependent on Galactic latitude (see Sect.~\ref{SubS_autodetect}, 
step 4). 
The mean cluster proper motions determined by the pipeline
for the 63 new clusters (see Table~\ref{Tmainpar}) also
confirm our initial rough estimates for the corresponding candidates.
The formal errors of the mean cluster proper motions are smaller than about
1~mas/yr because of the relatively large numbers of members per cluster.
The resulting total proper motions are relatively small. They lie in the 
range between 1~mas/yr and 13~mas/yr with a mean value of 
6.3~mas/yr. This is slightly larger than the average for
139 clusters from paper III (5.8~mas/yr) and the average for all 3006
clusters from paper II (5.5~mas/yr), but slightly smaller than the mean value
(7.0~mas/yr) for 231 MWSC open clusters with $|b|>$15$^{\circ}$ from paper II.

\begin{figure}
   \centering
   \includegraphics[width=7.0cm]{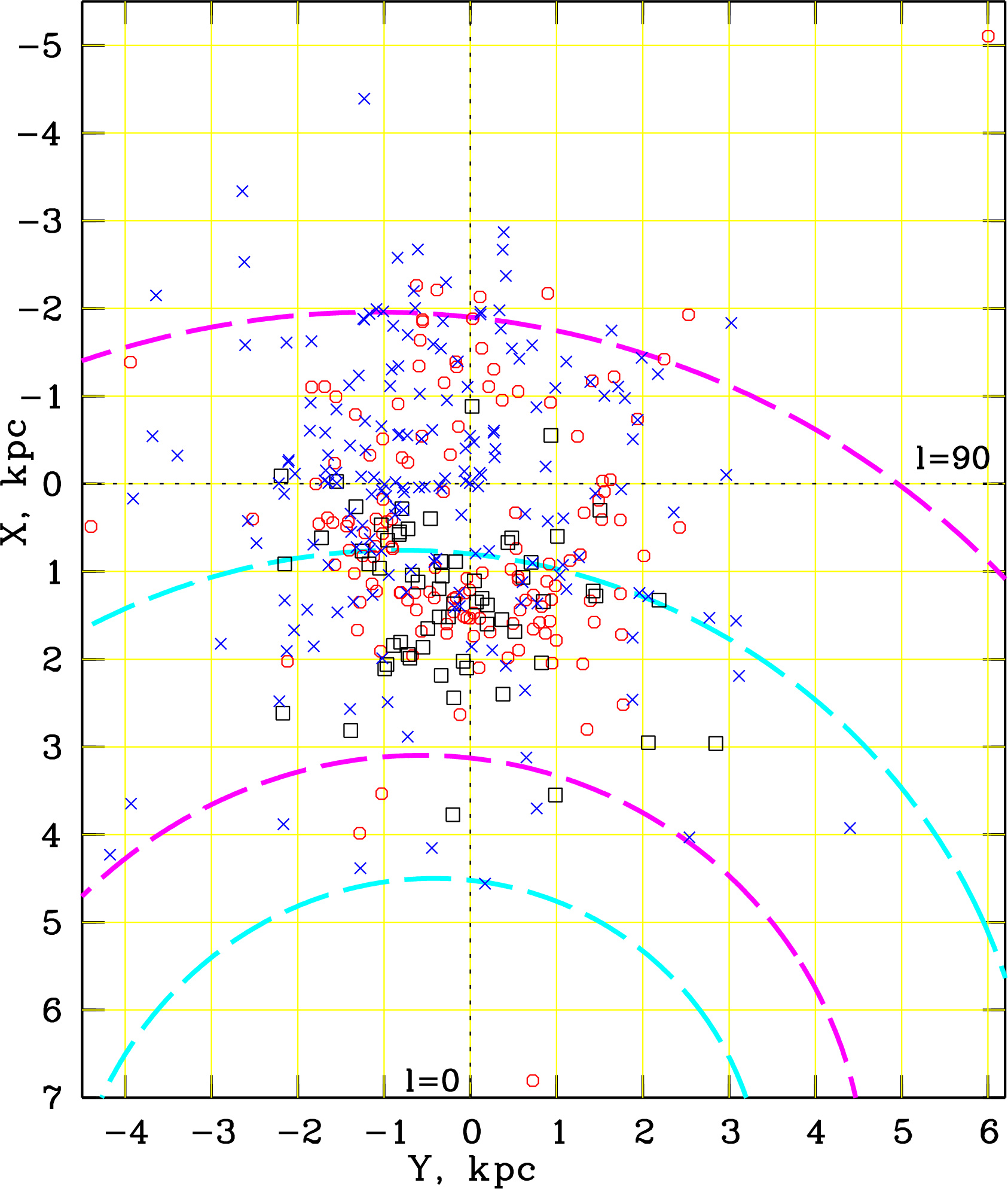}
      \caption{Distribution of clusters projected onto the Galactic $XY$-plane:
               blue crosses - 231 MWSC open clusters with $|b|>$15$^{\circ}$
                              from paper II,
               red open circles - 139 new open clusters from paper III,
               black open squares - 63 newly confirmed open clusters
               (this paper).
               The dashed spirals indicate the local spiral arms (magenta,
               first and third from top: Perseus arm; cyan, second and fourth
               from top: Sagittarius arm) as defined by COCD clusters
               (Piskunov et al.~\cite{piskunov06})
              }
         \label{Faiparixy}
   \end{figure}

An asymmetry between the numbers of newly found clusters in the outer and 
inner Galactic quadrants 
has already been observed 
for the discoveries
of paper III. This asymmetry is much stronger for our new 63 clusters
(see Fig.~\ref{Faiparixy}), from which only four lie in the outer
Galactic quadrants.

\begin{table*}%
\caption{Main parameters of 63 new open clusters.}
\label{Tmainpar}      
\centering                          
\fontsize{10pt}{0.75\baselineskip}\selectfont
\begin{tabular}{lccrrrrrrrr}        
\hline\hline                 
Name & RA (J2000)    & Dec (J2000)    & $l$ & $b$ & $r_2$ & $n_2$ & pmx      & pmy      & $d$  & $\log{t}$ \\
     & [h]   &[deg]   &[deg]&[deg]&[deg] &      & [mas/yr] & [mas/yr] & [pc] & [yr]\\
\hline 
MWSC\_4005 &  0.1910 & $-$85.480 & 303.850 & $-$31.573 &  0.165 &  42 &   $+$9.3 &   $-$1.5 &  1159 &  9.375 \\
MWSC\_4114 &  6.1310 & $-$78.220 & 289.714 & $-$28.731 &  0.240 &  60 &   $+$0.2 &   $+$5.9 &   962 &  9.400 \\
MWSC\_4116 &  6.3200 & $-$59.970 & 269.071 & $-$27.208 &  0.225 &  73 &   $-$0.9 &   $+$8.6 &  1743 &  9.125 \\
MWSC\_4119 &  7.1010 & $-$57.230 & 267.671 & $-$20.653 &  0.205 &  79 &   $-$2.8 &   $+$9.8 &  2347 &  8.985 \\
MWSC\_4131\tablefootmark{a,c} &  8.3215 & $+$42.035 & 178.664 & $+$33.536 &  0.220 &  37 &   $-$1.0 &   $-$7.7 &  1058 &  9.400 \\
MWSC\_4137 &  9.4365 & $-$79.740 & 294.396 & $-$20.461 &  0.180 &  83 &   $-$4.1 &   $+$8.8 &  1209 &  9.210 \\
MWSC\_4138 &  9.4520 & $-$73.500 & 289.628 & $-$16.163 &  0.185 & 103 &   $-$3.3 &   $+$6.5 &  1908 &  9.280 \\
MWSC\_4140 &  9.6100 & $-$77.480 & 293.037 & $-$18.529 &  0.225 & 112 &   $-$3.2 &   $+$3.3 &  2465 &  9.280 \\
MWSC\_4146\tablefootmark{b} & 11.0125 & $-$40.490 & 281.169 & $+$17.660 &  0.160 &  61 &   $-$9.4 &   $+$4.1 &  1419 &  9.100 \\
MWSC\_4158\tablefootmark{b} & 12.7410 & $-$38.310 & 301.420 & $+$24.544 &  0.195 &  80 &   $-$6.6 &   $-$3.6 &  1616 &  9.020 \\
MWSC\_4170 & 13.6135 & $-$39.185 & 312.447 & $+$22.850 &  0.220 & 132 &   $-$3.9 &   $-$0.5 &  1548 &  9.380 \\
MWSC\_4176 & 14.4550 & $-$85.420 & 304.948 & $-$22.925 &  0.280 & 155 &   $-$9.4 &   $+$0.1 &  1093 &  9.315 \\
MWSC\_4178 & 14.4605 & $-$44.945 & 320.272 & $+$14.667 &  0.180 &  94 &   $-$4.0 &   $-$3.6 &  3515 &  9.300 \\
MWSC\_4194 & 15.1315 & $-$17.425 & 343.417 & $+$34.512 &  0.320 &  69 &   $-$0.7 &   $-$1.0 &  2089 &  9.200 \\
MWSC\_4210 & 15.4640 & $-$36.520 & 334.835 & $+$16.487 &  0.205 & 168 &   $-$4.6 &   $-$4.3 &  2432 &  9.080 \\
MWSC\_4211 & 15.5265 & $-$37.490 & 334.869 & $+$15.262 &  0.240 & 190 &   $-$8.3 &   $-$4.0 &  2361 &  8.820 \\
MWSC\_4212 & 15.5530 & $-$30.335 & 339.699 & $+$20.760 &  0.225 &  99 &   $-$4.3 &   $-$5.0 &  2221 &  9.025 \\
MWSC\_4218 & 15.6740 & $-$33.475 & 338.909 & $+$17.343 &  0.250 & 128 &   $-$6.0 &   $-$7.1 &   993 &  9.100 \\
MWSC\_4219 & 15.7040 & $-$83.195 & 307.903 & $-$22.037 &  0.200 &  95 &   $-$9.3 &   $-$3.1 &  1606 &  9.100 \\
MWSC\_4225 & 15.8050 & $-$18.305 & 351.231 & $+$27.547 &  0.160 &  43 &   $-$3.6 &   $-$5.8 &  2495 &  8.700 \\
MWSC\_4228 & 15.9090 & $-$30.660 & 343.205 & $+$17.550 &  0.260 & 133 &   $-$4.7 &   $-$5.0 &  1312 &  8.740 \\
MWSC\_4229 & 15.9415 & $-$31.720 & 342.787 & $+$16.484 &  0.255 & 122 &   $-$4.8 &   $-$7.0 &  1148 &  9.015 \\
MWSC\_4248 & 16.2125 & $-$29.905 & 346.715 & $+$15.383 &  0.250 & 171 &   $-$1.1 &   $-$1.4 &  1618 &  8.900 \\
MWSC\_4265 & 16.4845 & $-$19.155 & 357.685 & $+$19.864 &  0.285 & 101 &   $+$0.2 &   $-$1.5 &  2153 &  8.600 \\
MWSC\_4267 & 16.4905 & $-$20.140 & 356.940 & $+$19.168 &  0.165 &  37 &   $-$3.2 &   $-$0.8 &  4001 &  8.250 \\
MWSC\_4286 & 16.6925 & $-$30.535 & 350.548 & $+$10.370 &  0.160 & 102 &   $+$0.1 &   $-$3.4 &  1564 &  8.910 \\
MWSC\_4287 & 16.7100 & $-$24.460 & 355.485 & $+$14.070 &  0.230 & 122 &   $-$4.5 &   $-$5.7 &  2524 &  7.920 \\
MWSC\_4288\tablefootmark{a} & 16.7365 &  $-$0.135 &  17.003 & $+$27.915 &  0.350 &  87 &   $-$0.1 &   $-$7.8 &  1994 &  9.015 \\
MWSC\_4289 & 16.7680 & $-$20.940 & 358.852 & $+$15.632 &  0.195 & 101 &   $-$0.8 &   $-$0.9 &  2183 &  9.325 \\
MWSC\_4290 & 16.7770 &  $-$9.195 &   8.956 & $+$22.483 &  0.165 &  44 &   $-$3.1 &   $-$5.3 &  2628 &  8.380 \\
MWSC\_4301 & 16.9890 & $+$18.360 &  38.180 & $+$32.723 &  0.295 &  68 &   $-$1.5 &   $-$7.0 &  1355 &  9.415 \\
MWSC\_4312 & 17.1085 & $-$19.445 &   3.030 & $+$12.665 &  0.220 & 133 &   $+$2.3 &   $-$3.9 &  1384 &  8.200 \\
MWSC\_4316 & 17.1745 & $+$51.570 &  78.642 & $+$36.379 &  0.225 &  35 &   $-$4.7 &   $+$1.6 &  1905 &  9.230 \\
MWSC\_4317\tablefootmark{a} & 17.1745 &  $+$1.265 &  21.991 & $+$22.930 &  0.190 &  61 &   $-$2.1 &   $-$0.2 &  2391 &  9.425 \\
MWSC\_4330 & 17.2840 & $-$15.080 &   8.182 & $+$13.033 &  0.175 & 126 &   $-$2.9 &   $-$4.4 &  1434 &  8.550 \\
MWSC\_4349 & 17.4505 & $-$57.525 & 333.768 & $-$12.271 &  0.150 &  84 &   $-$3.5 &   $-$7.7 &  3210 &  8.805 \\
MWSC\_4355 & 17.5100 & $-$23.935 &   2.437 &  $+$5.516 &  0.180 &  89 &   $+$1.4 &   $-$3.7 &  1113 &  8.680 \\
MWSC\_4375 & 17.8825 &  $+$9.515 &  34.944 & $+$17.280 &  0.200 & 106 &   $-$3.5 &   $-$2.8 &  3772 &  9.375 \\
MWSC\_4383 & 17.9790 & $+$23.050 &  48.688 & $+$21.423 &  0.210 &  71 &   $-$3.0 &   $-$6.9 &  2082 &  9.095 \\
MWSC\_4408 & 18.2585 & $-$60.215 & 334.203 & $-$19.064 &  0.165 &  82 &   $-$1.6 &   $-$8.0 &  2167 &  8.495 \\
MWSC\_4409\tablefootmark{a} & 18.2560 & $+$31.725 &  58.789 & $+$21.009 &  0.220 & 106 &   $+$0.3 &   $-$4.7 &  2740 &  8.850 \\
MWSC\_4410 & 18.2840 & $-$53.690 & 340.628 & $-$16.771 &  0.195 & 192 &   $-$4.6 &   $-$6.6 &  2202 &  8.900 \\
MWSC\_4429 & 18.5930 & $-$51.745 & 343.617 & $-$18.655 &  0.290 & 315 &   $-$0.5 &   $-$5.1 &  2051 &  9.045 \\
MWSC\_4463\tablefootmark{a} & 18.9820 & $-$64.265 & 331.551 & $-$25.042 &  0.210 & 125 &   $+$0.9 &   $-$5.4 &  1406 &  9.150 \\
MWSC\_4475 & 19.1360 & $-$30.230 &   6.977 & $-$16.590 &  0.165 &  96 &   $-$1.5 &   $-$3.3 &  1681 &  8.785 \\
MWSC\_4483 & 19.2855 & $-$68.525 & 327.128 & $-$27.432 &  0.235 & 113 &   $+$0.5 &   $-$5.8 &  1397 &  9.310 \\
MWSC\_4484 & 19.3720 & $-$25.010 &  13.271 & $-$17.519 &  0.210 & 146 &   $+$0.5 &   $-$5.2 &  1667 &  9.045 \\
MWSC\_4486 & 19.4000 & $-$60.725 & 335.952 & $-$27.341 &  0.180 &  65 &   $+$5.2 &   $-$7.7 &  2228 &  9.300 \\
MWSC\_4498\tablefootmark{a} & 19.5335 & $-$46.435 & 351.937 & $-$25.997 &  0.155 &  72 &   $+$1.5 &   $-$6.7 &  1537 &  9.360 \\
MWSC\_4500\tablefootmark{a} & 19.5560 & $-$33.235 &   5.964 & $-$22.657 &  0.150 &  68 &   $+$6.1 &   $-$1.2 &  1423 &  9.200 \\
MWSC\_4511 & 19.6940 & $-$24.455 &  15.539 & $-$21.390 &  0.170 &  72 &   $+$1.7 &   $-$4.2 &  3957 &  9.500 \\
MWSC\_4545\tablefootmark{a} & 20.1305 & $-$49.860 & 349.127 & $-$32.426 &  0.225 &  87 &   $+$1.9 &   $-$3.7 &  1070 &  9.310 \\
MWSC\_4547 & 20.2035 &  $+$1.525 &  43.847 & $-$17.151 &  0.165 &  50 &   $-$0.6 &   $-$2.0 &  4298 &  9.200 \\
MWSC\_4555\tablefootmark{a} & 20.2925 & $-$11.350 &  32.229 & $-$24.195 &  0.250 & 116 &   $+$7.1 &  $-$10.4 &  1740 &  9.505 \\
MWSC\_4561 & 20.3980 &  $+$5.900 &  49.340 & $-$17.490 &  0.185 &  63 &   $-$3.1 &  $-$11.7 &  1968 &  9.400 \\
MWSC\_4572\tablefootmark{a} & 20.5165 & $-$15.175 &  29.803 & $-$28.735 &  0.190 &  68 &   $+$3.6 &   $-$3.4 &  1402 &  9.340 \\
MWSC\_4602 & 21.0460 & $+$10.765 &  59.287 & $-$22.880 &  0.230 &  81 &   $+$3.1 &   $-$3.6 &  1274 &  8.785 \\
MWSC\_4606 & 21.1020 & $-$11.470 &  37.884 & $-$35.042 &  0.245 &  38 &   $+$2.5 &   $-$4.6 &   956 &  9.385 \\
MWSC\_4622 & 21.4340 & $-$17.750 &  32.917 & $-$41.939 &  0.240 &  31 &   $+$2.3 &   $-$8.5 &  1075 &  9.300 \\
MWSC\_4668\tablefootmark{a} & 22.9490 & $-$77.170 & 310.708 & $-$38.172 &  0.180 &  22 &   $+$8.2 &   $-$5.0 &   775 &  9.100 \\
MWSC\_4674 & 23.3690 & $+$84.010 & 120.487 & $+$21.563 &  0.210 &  63 &   $-$2.2 &   $+$0.1 &  1166 &  9.550 \\
MWSC\_4682 & 23.7230 & $-$82.960 & 305.405 & $-$33.835 &  0.190 &  58 &   $+$5.9 &   $+$0.3 &  1065 &  9.280 \\
MWSC\_4688 & 23.8650 & $-$86.720 & 303.906 & $-$30.291 &  0.185 &  57 &   $+$3.2 &   $-$5.0 &  1336 &  9.390 \\
\hline                        
\end{tabular}
\tablefoot{
\tablefoottext{a}{Sparse cluster}
\tablefoottext{b}{Cluster with poor radial density profile}
\tablefoottext{c}{Cluster remnant candidate}
}
\end{table*}

%

\section{Limitations of our cluster search and outlook}\label{Sect_limits}

The random and systematic errors of the stellar proper
motions in the available global catalogues (PPMXL and UCAC4) are larger than
or comparable in size to most of the mean cluster proper motions. Even after 
applying our HQ constraints, the quality assessment and comparison of our
chosen 2MAst\,HQb5 and UCAC4\,HQ samples forced us to use a rather large radius 
of 15~mas/yr for our circular proper motion bins. Compared to the accurate
2MASS photometry used as input in the search of paper III, the currently
available best proper motions did not allow us to reach a comparable
success rate in our proper motion-based cluster search.

There were much larger numbers of objects represented 
in our sky images 
with small proper motions compared to the sparsely
populated sky images with higher proper motions.
It may be 
worth concentrating on the latter, where the 
contamination with clusters of galaxies
should be much reduced, even if one considers the influence of the still
relatively large proper motion errors in the 2MAst\,HQb5 and UCAC4\,HQ samples. 
We also plan to develop a revised
method for detecting extended cluster candidates. Both aspects are
closely related to the question of missing nearby clusters.
These investigations will
be presented in a separate paper of the MWSC series.

The new clusters confirmed by the MWSC pipeline, which uses near-infrared
photometry, and the cluster parameters should be verified by independent
optical photometry. This is e.g. provided by the American Association of 
Variable Star Observers (AAVSO) Photometric all-sky survey
(APASS)\footnote{http://www.aavso.org/download-apass-data} for
a large part of the UCAC4 stars. Alves et al.~(\cite{alves12}) 
found some significant disagreements
between cluster parameters obtained from near-infrared 
and optical data. Other reduction pipelines for cluster
investigations like ASteCA (Automated Stellar Cluster Analysis) by
Perren et al.~(\cite{perren15}) 
or the program of Popova~(\cite{popova13})
can be applied for the verification of our 63 new clusters.
The use of the UCAC4 proper motions (as e.g. by Dias et al.~\cite{dias14})
may also lead to some changes in the cluster membership and mean cluster
proper motions. The latter may differ by a few mas/yr according to our
comparison of 2MAst\,HQb5 and UCAC4\,HQ proper motions over the sky.

An enormous quality jump in the input data for a search for Galactic clusters
can be expected from the Gaia mission. This concerns especially the distance
measurements, which will be first provided for the same about 2.5 million 
bright stars (Michalik et al.~\cite{michalik15}), which we used in our first 
cluster search, the COCD extension (Kharchenko et al.\cite{kharchenko05b}).
While at that time only the small Hipparcos subsample was equipped with
parallaxes, we expect
the first Gaia data release to provide
about 10-20 times higher accuracy for the parallaxes and proper motions of 
all these 2.5 million stars, respectively. The 
second Gaia data release will 
then deliver proper motions and parallaxes for a number
of objects with single-star behaviour that is probably at least as large as
the number of objects in our 2MAst\,HQ and UCAC4\,HQ samples, where the
parallaxes will be available for the first time, whereas the proper motions
are expected to be 10-100 times more accurate. 
Therefore, a proper motion-based search for new clusters in the Gaia
data will be possible with very small circular proper motion bins 
(with radii of the order of 1~mas/yr). 
In addition, all these stars
will have optical photometry, and a brighter subset will have radial velocity 
measurements. Sky distributions of Gaia subsamples selected not only by
proper motions but also by distances and radial velocties can be investigated
in much more detail (on a fine grid of these parameters) to first investigate
possible small systematic errors over the sky and then to search for Galactic 
clusters. In addition to the much higher accuracies of the measured stellar 
parameters, the high spatial resolution of Gaia is an advantage
for detecting star clusters in crowded fields.

%
\begin{acknowledgements}
This study was supported by DFG grant RO 528/10-1, and  by Sonderforschungsbereich SFB 881 "The Milky Way System" (subproject B5) of the German Research Foundation (DFG).
This research has made use of SAOImage DS9, developed 
by Smithsonian Astrophysical Observatory,
and of data products from the 2MASS, a joint project
of the University of Massachusetts and the Infrared Processing and Analysis
Center/California Institute of Technology, funded by the 
National Aeronautics and Space Administration and the
National Science Foundation.
We thank the anonymous referee for her/his helpful comments.
\end{acknowledgements}


\end{document}